\documentclass{article}
\usepackage{lipsum}
\usepackage{authblk}
\usepackage[english]{babel}
\usepackage{amsfonts}%pour police des  symboles mathematiques (gras ,italique, ....)
\usepackage{amsmath} %améliorer la structure de l'information et la sortie imprimée de documents contenant des formules mathématiques
\usepackage[top=2cm, bottom=2cm, left=2cm, right=2cm]{geometry}% format du papier
\usepackage{fancyhdr} %style du page
\usepackage{multicol} % deux colonne dans l'article
\usepackage{epsfig} %extention   eps
\usepackage{graphics} % pour inserer correctement une image
\usepackage{epsfig}
\usepackage{lipsum}
\usepackage{fancyhdr}
\usepackage{ulem}
\usepackage{xcolor}
\usepackage{booktabs}
\usepackage{subcaption} 
\usepackage[export]{adjustbox} 
\usepackage{cleveref}
\usepackage{wrapfig} 

\renewenvironment{abstract}{
	
	\hfill\begin{minipage}{0.95\textwidth}
		\rule{\textwidth}{1pt}}
		{\par\noindent\rule{\textwidth}{1pt}\end{minipage}
	}

\begin{document}
%
%title and author details
\title{\textbf{Engineering entanglement and teleportation via solving Lindblad master equation}}
\author[1]{\textbf{K. El Anouz}}
\author[1,2]{\textbf{A. El Allati}}
\author[3]{\textbf{F. Saif}}
\affil[1]{\small Laboratory of R\&D in Engineering Sciences, Faculty of Sciences and Techniques Al-Hoceima, Abdelmalek Essaadi University, T\'{e}touan,
	Morocco}
\affil[2]{\small  The Abdus Salam International Center for Theoretical Physics,\\
Strada Costiera 11, Miramare-Trieste, Italy}
\affil[3]{\small  Department of Electronics, Quaid-i-Azam University, 45320 Islamabad, Pakistan}
\maketitle
 \begin{center}
	\textbf{Abstract}
\end{center}
\begin{abstract}
 An experimentally realizable model based on the interaction between  an excited two-level atom and a radiation field inside two quantum electrodynamics cavities is proposed. It consists of sending an excited two-level atom through two serial cavities which  contain the radiation field. Hence, the Lindblad master equation described the reduced density matrix of the joint-joint field system inside the cavities is exactly solved in Markovian and non-Markovian regimes. However, the rate of entanglement inherent in the total field-field system are evaluated using various witnesses of entanglement such as concurrence, logarithmic negativity and quantum discord. Moreover, the non-classicality by means of  negativity volume and  Wigner function is discussed. Finally, two schemes of quantum teleportation are suggested.
\end{abstract}

\section{Introduction}

The theory open quantum system plays a substantial  role in the last decade. It is considered as a brilliant theory and one of the important theories until now \cite{1,2}. Indeed, it is at the heart many concepts needed to understand and manipulate the interaction between physical a system and its surrounding. Actuality, the core idea underlying the theory o open systems is  to separate a global system containing such several subsystems  into two parts: the most important degrees of freedom, which constitute what is called open quantum system, are treated explicitly; the other degrees of freedom, which belong to the environment, appear only implicitly, where in general the open system can exchange energy with its environment. However, this theory leads to perform many tasks in different disciplines including quantum information processing \cite{3,4,5,6}, quantum optics \cite{5}, atom-cavity interactions \cite{1,7}. On the other hand, the interaction between any open system and its surrounding environment gives rise to the dissipation and decoherence phenomena \cite{1,8}. The dissipative dynamics methods give rise to a fundamental equation governed the interaction between the open quantum system and environment. This equation is often called Master Equation \cite{8a}. In this context, many works are investigated; in both theoretical and experimental levels to explore the dynamics of open quantum systems \cite{9,10,11}. Moreover, many methods have been investigated in order to derive the fundamental equations described the interaction between the open system and its environment. Born-Markov approximation and projection operator method are the most successful assumptions used to describe the Markovian and non-Markovian master equations, respectively \cite{1,3}. Roughly speaking, the open system is always correlated to its environment where they cannot be seen as two separable parts even the distance between them is large. Hence, several  measures are introduced in the literature to examine the correlations between quantum systems.\\

In 1935, Erwine Schr\"odinger introduced entanglement as a kind of quantum correlation  which is the physical phenomenon described the interaction between two subsystems \cite{12}. In general, quantum entanglement reflects the phenomenon in which two arbitrary quantum systems removed from each other construct a single system in such a way if we generate an action on one of them allows to distribute the other one. In this inspiration,  various entangled states have been studied such as NOON states, GHZ states, W states and cluster states etc.~\cite{s1,s2,s3}. However, many works paid attention to study entanglement. For example Franco $et$ $al.$  investigated the dynamics of quantum correlations for a bipartite system coupled to a non-Markovian environment \cite{13}. Moreover, Coladangelo presented an avenue for device-independent certification of maximally entangled states at arbitrary local dimension \cite{14}. Moreover, K. Berrada $et$ $al.$ studied the quantum correlation dynamics of a two identical qubits interacted with a bosonic reservoir under non-Markovian regime \cite{15}.  However,  the entanglement phenomenon can be quantified by the rate of entanglement inherent in a quantum physical system by means of the so called $entanglement$ $measures$ \cite{18,19,16,17}. Vidal and Werner showed that the logarithmic negativity represents a good entanglement measure \cite{20}. Additionally, concurrence is the most popular quantitative measure of entanglement \cite{21}. Moreover both measures are equal to zero if and only if the states are separable and equal to one if they are maximally entangled. Extensively, the quantum discord is one of the most success quantifier of quantum correlations \cite{22}. It is defined as the difference between classical correlation and quantum mutual information. For pure sates, the quantum correlation is equivalent to the entanglement entropy \cite{3}. However, for mixture Bell states, it takes the maximum value which is one. Unluckily for two mixed states it is so complicated to investigate the quantum discord dynamics. Recently, the concept of negativity volume based on the Wigner function serves as an identifier of entanglement, purity and non-classicality of quantum states \cite{Wigner 1, Wigner 2}.\\

In this paper, we assume that a two-level atom passes successively through two serial cavities $A$ and $B$ each one of them is described by the coherent state \cite{1,6}, namely $|\alpha\rangle $ and $|\beta\rangle $, respectively. However, by solving exactly the master equation for the reduced density matrix of the field cavities system, we investigate the quantum correlation of the joint field-field state by means of concurrence, logarithmic negativity and quantum discord. Moreover we study the evolution of Wigner function as well as the negativity volume of the total field-field density matrix to display its non-classicality. Finally, the entangled joint field-field state is used as a quantum channel in order to implement two schemes of quantum teleportation protocol \cite{23,24,25}. By controlling the cavity field parameters we examine the dynamics of  the teleported entanglement measures as well as fidelity of the teleported state. Our results show that the entanglement degrees as well as the teleported entanglement degrees in the non-Markovian regimes vanish fast comparing to those obtained in the Markovian approach, where the sudden-death-time phenomenon depends in general on  the cavity-field parameters.\\

Our paper is organized as follows: in the next section our proposed model is presented. In sections $3$ and $4$, we introduce the general definitions and preliminaries of logarithmic negativity, concurrence and quantum discord as well as the volume negativity and Wigner function, respectively. In section $5$, we examine the evolution of concurrence, quantum discord and logarithmic negativity using two various initial entangled states, namely (EPR) and (NOON) states. In section $6$ as an application, two schemes of quantum teleportation are investigated using the obtained results. We conclude our discussion by a summary and some future perspectives in section $7$.

\section{Proposed model}

Assume that a two-level atom of the upper and lower levels $|k\rangle $ and $|l \rangle $, respectively is initially prepared in the excited state $|k\rangle $. This atom crosses successively two serial cavities $A$ and $B$ in such a way a radiation field of frequency $\nu$ is generated. However, let suppose that the radiation fields inside the cavities $A$ and $B$ have the same frequency $\nu_k$ of $k$ modes (see Fig. (1)). In the interaction picture and by using the rotating-wave approximation, the interaction Hamiltonian takes the following form \cite{29a, 29}
\begin{equation}\label{H}
H_I(t)=\hbar \sum_{k}\sum_{j=A,B}[g_k^{j} b_k^{j^{\dagger}} a^j e^{-i(\nu-\nu_k)t}+g_k^{*j} a^{j^{\dagger}} b_k^{j} e^{i(\nu-\nu_k)t}],
\end{equation}
 \begin{figure}[h!]
	\begin{center}
		\includegraphics[scale=1]{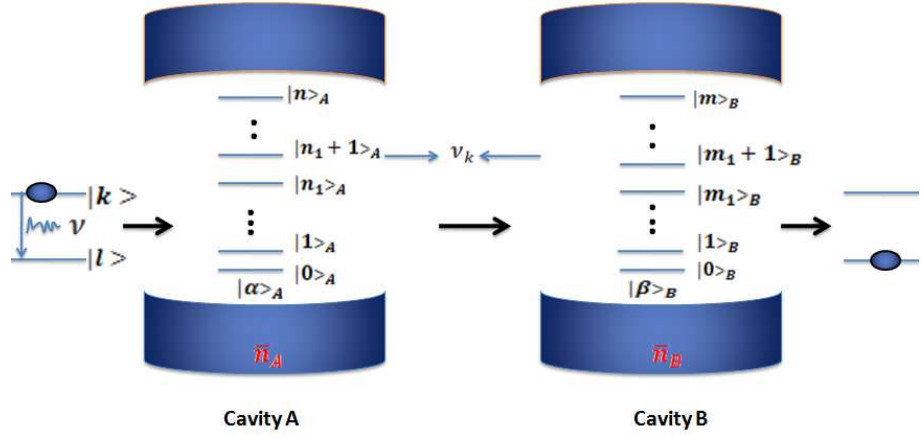}
		\caption{The scheme for the interacted  joint field-field state; a generated field of frequency $\nu$ arrived from an excited two-level atom passes successively through two serial cavities $A$ and $B$, respectively of frequency $\nu_k$.}
	\end{center}
\end{figure}

where $a^{A (B)}$ and $a^{A (B)^\dagger}$ are the annihilation and creation operators of the electromagnetic field generated from the excited atom, respectively. However, $b^{A (B)}_{k}$ and $b^{A (B)^\dagger}_{k}$ denote the annihilation and creation operators of the electromagnetic field of frequency $\nu_k$ inside the cavities. Moreover,  $g_{k} ^{A(B)}$ represents the constant coupling. On the other hand, the evolution of the joint field-field inside the cavities is governed by the Lindblad master equation of the form \cite{1, 29},
\begin{equation}\label{Lindblad M.E}
\dot{\rho}_S^{L}(t)=\sum_{j=A,B} \Big[-\frac{\gamma ^{j}(t)}{2} (\bar{n}_{j}+1)[a_{j}^{\dagger} a_j \rho_S(t)+\rho_S(t) a_{j}^{\dagger} a_j-2 a_j\rho_S(t) a_j^{\dagger}]-\frac{\gamma ^{j}(t)}{2} \bar{n}_{j}[  a_j a_{j}^{\dagger} \rho_S(t)+\rho_S(t)  a_j a_{j}^{\dagger}-2 a_j^{\dagger}\rho_S(t) a_j]\Big],
\end{equation}

where $\gamma^j(t)$ ($j=A, B $) denotes the damping rate of the cavities, while $\bar{n}_A$ and $\bar{n}_B$ are the average numbers. Moreover, $\bar{n}_j$ is connected directly to the temperature degree $T$ as  bellow \cite{1}
\begin{equation}
\bar{n}_j=\frac{1}{e^{\hbar\nu/k_{B}T}-1}.
\end{equation}
As is clear, due to the vacuum fluctuation, i.e, $T=0$, the average numbers is  vanished. In general, the damping rate $\gamma^j(t)$ reflects the memory effects arrived from the interaction between the open system and its surrounding environment. Under Markov approximation, the memory effects are short-lived, i.e. the rate $\gamma^j(t)$ can be replaced by $\gamma^j_M$. Hence, the Markov master equation can rewritten as follows \cite{1, lind}
\begin{equation}\label{Markovian M.E}
\dot{\rho}_S^{M}(t)=\sum_{j=A,B} \Big[-\frac{\gamma ^{j}_M}{2} (\bar{n}_{j}+1)[a_{j}^{\dagger} a_j \rho_S(t)+\rho_S(t) a_{j}^{\dagger} a_j-2 a_j\rho_S(t) a_j^{\dagger}]- \frac{\gamma ^{j}_M}{2} \bar{n}_{j}[  a_j a_{j}^{\dagger} \rho_S(t)+\rho(t)  a_j a_{j}^{\dagger}-2 a_j^{\dagger}\rho_S(t) a_j]\Big].
 \end{equation}
However, in the non-Markovian dynamics, the damping rate depends on the spectral structure of the environment \cite{1}. In this model, we suppose that the damping rate  is evaluated for an Ohmic reservoir by means of Lorentz-Drude cut-off function as bellow \cite{26}
\begin{eqnarray}\label{gamma}
\gamma(t)&=&\int_{0}^{t} ds \mu(s),\nonumber\\
&=& 2 \int_{0}^{t} ds \int_{0}^{\infty} d\omega J(\omega) \sin(\omega s),
\end{eqnarray}
where $\mu(s)$ and $J(\omega)$ denote the dissipation kernel and the Ohmic spectral density, respectively. Moreover,the Ohmic spectral density is collapses to be,
\begin{equation}\label{J(omega)}
J(\omega)=\frac{2\omega}{\pi} \frac{\omega_{c}^2}{\omega_c^2+\omega^2},
\end{equation}
where $\omega$ and  $\omega_c$ are the bath and the environment spectrum  cut-off frequencies, respectively. Using Eqs.(\ref{Lindblad M.E}), (\ref{gamma}) and (\ref{J(omega)}), the non-Markovian master equation turns out to be \cite{1,15}
  \begin{equation}\label{non-Markovian M.E}
  \dot{\rho}_S^{NM}(t)=\sum_{j=A,B} \Big[-\frac{\Gamma (t)}{2} (\bar{n}_{j}+1)[a_{j}^{\dagger} a_j \rho_s(t) -2 a_j\rho_s(t) a_j^{\dagger} +\rho_s(t) a_{j}^{\dagger} a_j]- \frac{\Gamma (t)}{2} \bar{n}_{j}[  a_j a_{j}^{\dagger} \rho_s(t) -2 a_j^{\dagger}\rho_s(t) a_j +\rho_s(t)  a_j a_{j}^{\dagger}]\Big],
  \end{equation}
where
\begin{equation}\label{Gamma}
\Gamma(t)=\frac{8r^2}{1+r^2}[\omega_0 t+\frac{r-1}{1+r^2} e^{r\omega_0 t} \sin(\omega_0 t)+\frac{2r}{1+r^2}(e^{r\omega_0 t} \cos(\omega_0 t)-1) ], \quad r=\frac{\omega_c}{\omega_0},
\end{equation}
$\omega_0$ describes the frequency-independent damping constant. Let assume that the initial joint field-field system inside the cavities $A$ and $B$ is given by the following tensor product:
\begin{equation}\label{initi}
|\phi\rangle_{AB}(0)= |\alpha\rangle_A \otimes |\beta\rangle_B,
\end{equation}
where $ |\alpha\rangle_A$ and $ |\beta\rangle_B$ denote the coherent states describing the field inside the cavities $A$ and $B$, respectively. They are expressed as bellow \cite{jh}
\begin{equation}
|\alpha \rangle_A =\exp(-\frac{\bar{n}_A^{'}}{2}) \sum_{n=0}^{\infty}\sqrt{\frac{\bar{n}_A^{'^n}}{n!}} |n\rangle\quad,\quad
|\beta \rangle_B =\exp(-\frac{\bar{n}_B^{'}}{2}) \sum_{m=0}^{\infty}\sqrt{\frac{\bar{n}_B^{'^m}}{m!}} |m \rangle,
\end{equation}
where $\bar{n}_A^{'}$ and $\bar{n}_B^{'}$ are the mean photon numbers, while $|n\rangle$ and $|m\rangle$ are  the cavity Fock states. For sake of simplicity we suppose that  $\bar{n}_A^{'}=\bar{n}_B^{'}=\bar{n}^{'}$, then the initial state $|\phi\rangle_{AB}(0)$ of Eq.(\ref{initi}) is decomposed in Fock basis as
\begin{equation}\label{total initial state}
|\phi\rangle_{AB}(0)= a |n_1,m_1\rangle+b|n_1,m_1+1 \rangle +c|n_1+1, m_1\rangle +d |n_1+1,m_1+1\rangle,
 \end{equation}
where the different probability amplitudes turn out to be,
 \begin{eqnarray}
 a&=& \frac{1}{\sqrt{2}} e^{-\bar{n}^{'}} \sqrt{\frac{\bar{n^{'}}^{n_1+m_1}}{(n_1m_1)!}}\quad,\quad\qquad b= \frac{1}{\sqrt{2}} e^{-\bar{n}^{'}} \sqrt{\frac{\bar{n^{'}}^{n_1+m_1+1}}{[n_1(m_1+1)]!}},\nonumber\\
 c&=& \frac{1}{\sqrt{2}} e^{-\bar{n}^{'}} \sqrt{\frac{\bar{n^{'}}^{m_1(n_1+1)}}{[m_1(n_1+1)]!}},\quad\quad d= \frac{1}{\sqrt{2}} e^{-\bar{n}^{'}} \sqrt{\frac{\bar{n^{'}}^{(m_1+1)(n_1+1)}}{[(m_1+1)(n_1+1)]!}},
 \end{eqnarray}
 where $|a|^2+|b|^2+|c|^2+|d|^2=1$. Moreover, the states $|n_1\rangle $ and $|m_1\rangle $ are chosen arbitrary from the Fock basis (see Fig. (1)). Building on that foundation, one can exactly solve the Markovian and non-Markovian master equations defined in Eqs.(\ref{Markovian M.E}) and (\ref{non-Markovian M.E}), respectively. Indeed, the reduced density matrix of the joint field-field system in the basis $\{ |n_1,m_1\rangle, |n_1,m_1+1 \rangle, |n_1+1,m_1\rangle, |n_1+1,m_1+1\rangle \}$ is collapses to be, 
	\begin{equation}\label{rho}
\rho(t)=\begin{pmatrix}
\rho_{11}(t)  & \rho_{12}(t) & \rho_{13}(t) & \rho_{14}(t) \\
\rho_{21}(t) & \rho_{22}(t) & \rho_{23}(t) & \rho_{24}(t) \\
\rho_{31}(t) & \rho_{23}^{*}(t) & \rho_{33}(t) & \rho_{34}(t) \\
\rho_{14}^{*}(t) & 	\rho_{24}(t)  & 	\rho_{34}(t)  & \rho_{44}(t)
\end{pmatrix},
\end{equation}
where, the elements $\rho_{11} (t), ..., \rho_{44} (t)$ are calculated in appendix $B$.

\section{Different degrees of quantum correlations}
In  this section, we call back different witnesses of quantum correlations commonly used throughout this work to quantify the separability of the joint field-field state.

 	\begin{description}
 		\item[a/] \textbf{Negativity and logarithmic negativity}
 		
Given any operator $\rho$, the  trace norm, i.e, the singular values sum of $\rho$ defined as $||\rho||_1=Tr|\rho|=Tr\sqrt{\rho^{\dagger} \rho}$ \cite{27,28} gives rise to the following definition of  negativity \cite{29}
\begin{equation}\label{neg}
 \mathcal{N}(\rho)=\frac{||\rho^{T_{B}}||_1-1}{2},
\end{equation}
where $\rho^{T_{B}}$ denotes the partial transpose of $\rho$ in $d_1 \otimes d_2$ dimension. For maximally entangled states, the above formula in Eq. (\ref{neg}) ensures that the negativity coincides with the entropy of entanglement . Moreover, for any bipartite state the logarithmic negativity ($LN$) is defined as bellow \cite{ln}
 		\begin{equation}\label{LN}
 		LN(\rho)=\log_{2}||\rho^{T_{B}}||_1 \qquad\hbox{with}\quad ||\rho^{T_{B}}||_1=1+2|\sum_{i}\lambda_i|,
 		\end{equation}
where $\lambda_i$ denote the negative eigenvalues of $\rho^{T_{B}}$.
 		\item[b/] \textbf{Concurrence}

Concurrence can be known as the most popular and successful witness of entanglement used in quantum information theory, thanks to it's simplicity and efficiency. The concurrence takes the following compact form \cite{21}
\begin{equation}\label{con}
C(\rho)=\max(0,\sqrt{\lambda_{1}}-\sqrt{\lambda_{2}}-\sqrt{\lambda_{3}}-\sqrt{\lambda_{4}}),
 	\end{equation}
where $\lambda_{k}$ $(k=1...4)$ are the eigenvalues of the matrix
$\rho (\sigma_{y}^{A}\otimes \sigma_{y}^{B})\rho^{\ast}(\sigma_{y}^{A}\otimes \sigma_{y}^{B})$ and  $\sqrt{\lambda_{1}}\geq\sqrt{\lambda_{2}}\geq\sqrt{\lambda_{3}}\geq\sqrt{\lambda_{4}}$. Moreover $\sigma_{y}^{\alpha}$ denotes the second Pauli matrix for a qubit $\alpha$, while $\rho$ defines the reduced density operator of the total system. In general,
$0\le C(\rho) \le 1$. Indeed,  if $C(\rho)=0$, then $\rho$ becomes a separable state, while if $C(\rho)=1$ then the state is maximally entangled.
 	
\item[c/] \textbf{Quantum Discord}

Quantum discord is one of the most useful quantifier of quantum correlations in quantum systems. The
main idea behind defining the quantum discord is to show that even a separable state may contain quantum correlations by means of its total and classical correlations. Indeed, quantum discord is defined as the difference between quantum mutual information and classical correlation. For a bipartite quantum state $\rho^{AB}$, the quantum discord is expressed as bellow \cite{30,31}
 	\begin{equation}\label{discord}
 	QD(\rho^{AB})=I(\rho^{AB})-CC(\rho^{AB}),
 	\end{equation}
 where $ 	I(\rho^{AB})=S(\rho^{A})+S(\rho^{B})-S(\rho^{AB}).$ denotes the quantum mutual information and 
$S(\rho^{AB})=-Tr(\rho^{AB}\log_{2}\rho^{AB})$ is the von Neumann entropy of the total system. However,  $CC(\rho^{AB})$ represents  the classical correlation of the system $AB$ as bellow
 	\begin{equation}\label{cc}
 	CC(\rho^{AB})=S(\rho^{A})- \min_{B_{m}} \sum_{m} P_{m} S(\rho^{A}_{m}),
 	\end{equation}
The minimum indicated in Eq. (\ref{cc}) is taken over the orthogonal projectors set $ \{ B_m\} $. Moreover $\sum_{m} P_{m} S(\rho_{A}^{m})$ defines the quantum conditional entropy of the outcome post measurement state $\rho_{A}^{m}=\frac{1}{P_m}Tr_{B}[(I\otimes B_{m}) \rho_{AB}(I \otimes B_{m})]$ and the probability $P_{m}=Tr_{AB}[(I\otimes B_{m})\rho_{AB}(I\otimes B_{m})]$. Hence the explicit formula of quantum discord is defined as follows \cite{33}
\begin{equation}\label{QD}
 	QD(\rho_{AB})=\min (Q_{1},Q_{2})\hbox{ with } Q_{i}=H(\rho_{11}+\rho_{33})+\sum_{j=1}^{4}\lambda_{i} \log_{2}\lambda_{j}+D_{i}, \quad (i=1,2),
\end{equation}
where $\lambda_{j}$ are the eigenvalues of $\rho_{AB}$ and $H()$ is the entropy of Shannon. Moreover, $D_1$ and $D_2$ are given by the following formulas:
  	\begin{equation}
 	D_{1}=H(s), \quad D_{2}=-\sum_{j=1}^{4}\rho_{jj}-H(\rho_{11}+\rho_{33}),\quad\hbox{with}\quad
 	s=\frac{1}{2}(1+\sqrt{[1-2(\rho_{33}+\rho_{44})]^{2}+4(|\rho_{14}|+|\rho_{23}|)^2}).
 	\end{equation}
\end{description}

\section{Wigner function and negativity volume as an entanglement measure }
In  quantum mechanics, the  Wigner  function  supports  all   probability  distributions because it is non  singular and real. Indeed, the   Wigner distribution function has been used in many physical systems due to its symmetric properties. Here we present the negativity volume by means of Winger function, which depicts non classicality criterion as a measure of entanglement. The joint Wigner function of the field inside the cavities is defined as bellow \cite{Wigner 1}
\begin{eqnarray}\label{Wigner 1}
W(\alpha, \beta)&=&\frac{4}{\pi^2}<P(\alpha, \beta)>\nonumber\\
&=& \frac{4}{\pi^2} Tr[\rho D_{\alpha}  D_{\beta} P  D_{\beta}^{\dagger}  D_{\alpha}^{\dagger}],
\end{eqnarray}
where $P=P_A P_B= e^{i \pi a^{\dagger} a} e^{i \pi b^{\dagger} b} $ is the total parity and $D_{\alpha (\beta)}$ denote the displacement operator for the cavities $A$ and $B$, respectively. The linearity of Wigner function allows to simplify the expression given in Eq.(\ref{Wigner 1}) as follows \cite{Wigner 1}
\begin{eqnarray}
W(\rho, \alpha, \beta)&=& \frac{4}{\pi^2} Tr[\rho D P_A P_B  D^{\dagger}]\,,\nonumber\\
&=& \frac{4}{\pi^2} \sum_{ijkl} \rho_{ijkl} \langle kl| D P_A P_B  D^{\dagger} | ij \rangle \,, \nonumber\\
&=& \frac{4}{\pi^2} \sum_{ijkl} \rho_{ijkl} K_{ki}^{A} K_{lj}^{B}.
\end{eqnarray}
In our case, the matrix elements $K_{m m^{'}}^{A(B)}$ of Eq. (\ref{rho}) are collapse to be,
\begin{eqnarray}
K_{m m^{'}}^{i}&=& \langle m| D P D^{\dagger} |m^{'}\rangle\nonumber\\
&=& e^{-|\alpha|^2} (-1)^m (2|\alpha|)^{m^{'}-m} \sqrt{\frac{m}{m^{'}!}} L_m^{m^{'}-m} (|\alpha|),
\end{eqnarray}

where $i=(A, B)$. Moreover, $L_m^{m^{'}-m} (|\alpha|)$ gives the generalized Laguerre polynomial \cite{Wigner 1}. However, in order to identify the non-classicality of a quantum state, the so called negativity volume by means of Wigner function may is taken into account. It is expressed as follows,
\begin{equation}
V=\frac{1}{2} (\int |W|-W) d\Omega,
\end{equation}
where the normalization condition  $\int W d\Omega=1$ gives rise to:
\begin{equation}\label{volu}
V=\frac{1}{2} (\int |W|-1) d\Omega.
\end{equation}
The main motivation behind using the negativity volume is given in the possibility to exploit the total information encoded in a quantum state. For example, it can be considered as a good quantifier of entanglement. However, the Wigner function can be negative even for the classical states which provides a sufficient condition for the  non-classicality  of a quantum  state. In the next section, we shall use the joint field-field state in  Eq.(\ref{total initial state}) to solve the Lindblad master equations given in Eqs.  (\ref{Markovian M.E}) and (\ref{non-Markovian M.E}). The second topic consists of quantifying the entanglement rate inherent in the resulting field-field state by means of different entanglement degrees already depicted. Based on this, two schemes of quantum teleportation will be proposed. 
\section{Numerical results and discussion}
unlucky the quantification of the entanglement degrees of the total field-field density operator founded in Eq.(\ref{rho}) seems to be difficult. For sake of simplicity, we shall prepare initially the joint field-field state inside the cavities A and B in two various initial states, namely the $EPR$ and $NOON$ states instead the total initial state (\ref{total initial state})
 \begin{enumerate}
 	\item \underline{First case}: In this case we assume the total field-field state is initially prepared in $EPR$ state, i.e, $|\phi\rangle _{AB} ^1(0)= a |n_1, m_1\rangle+d |n_1+1, m_1+1\rangle$, that is, $b=c=0$ in Eq.(\ref{total initial state}). However, using the results in appendix $B$, one can simply rewrite the state in Eq.(\ref{rho}) as bellow,
 	\begin{equation}\label{rho1}
 	\rho^{1}(t)=\begin{pmatrix}
 	\rho_{11}(t)  & 0 & 0 & \rho_{14}(t) \\
 	0 & \rho_{22}(t) & 0 & 0 \\
 	0 & 0 & \rho_{33}(t) & 0 \\
 	\rho_{14}^{*}(t) & 0 & 0 & \rho_{44}(t)
 	\end{pmatrix},
 	\end{equation}
where the above elements $\rho_{ii}$ and $\rho_{ij}$ , $i \neq j$ are given in appendix B. On the other hand, the different measures of quantum correlations already presented in Sec. (3) are collapse to be,
\begin{eqnarray}\label{degree of entanglement 1}
 	C_1(t)&=&\max(0,2[|\rho_{14}(t)|-\sqrt{\rho_{22}(t) \rho_{33} (t)}]),\nonumber\\
 LN_1(t)&=&\max(0,\log_{2}[1-\rho_{22}(t)-\rho_{33}(t)+\sqrt{[\rho_{22}(t)-\rho_{33}(t)]^2+4|\rho_{14}(t)|^2}]),\nonumber\\
 QD_1(t)&=&\min(Q_1^1,Q_2^1),
 \end{eqnarray}
 where
 \begin{eqnarray}
 Q_1^1&=&H(\rho_{11}(t)+\rho_{33}(t))+ \sum_{i=1}^{4} \lambda_i \log_{2}\lambda_i+H(\frac{1}{2}(1+\sqrt{[1-2 (\rho_{33}(t) \rho_{44}(t)]^2+4|\rho_{14}(t)|^2 })),\nonumber\\
 Q_2^2&=&\sum_{i=1}^{4} \lambda_i \log_{2}\lambda_i+\sum_{i=1}^{4} \rho_{ii}.
 \end{eqnarray}
\begin{figure}[h!]
	\centering
	
	\begin{subfigure}[t]{0.03\textwidth}
		\textbf{(a)}
	\end{subfigure}
	\begin{subfigure}[t]{0.4\textwidth}
		\includegraphics[width=\linewidth,valign=t]{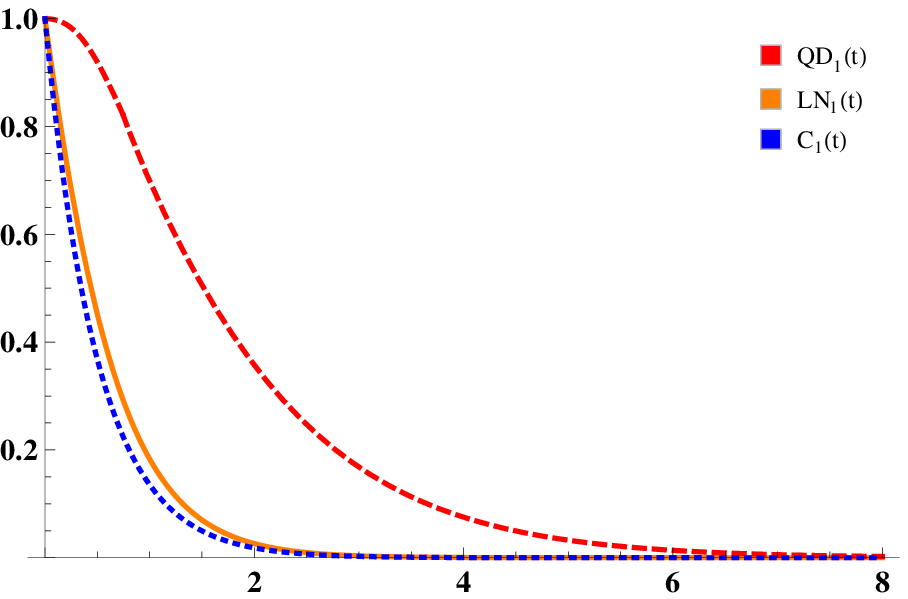}
	\put(2,-115){$\gamma_M t$}~~\quad
	\end{subfigure}
\hfill
	\begin{subfigure}[t]{0.03\textwidth}
		\textbf{(b)}  
	\end{subfigure}
	\begin{subfigure}[t]{0.4\textwidth}
	\includegraphics[width=\linewidth,valign=t]{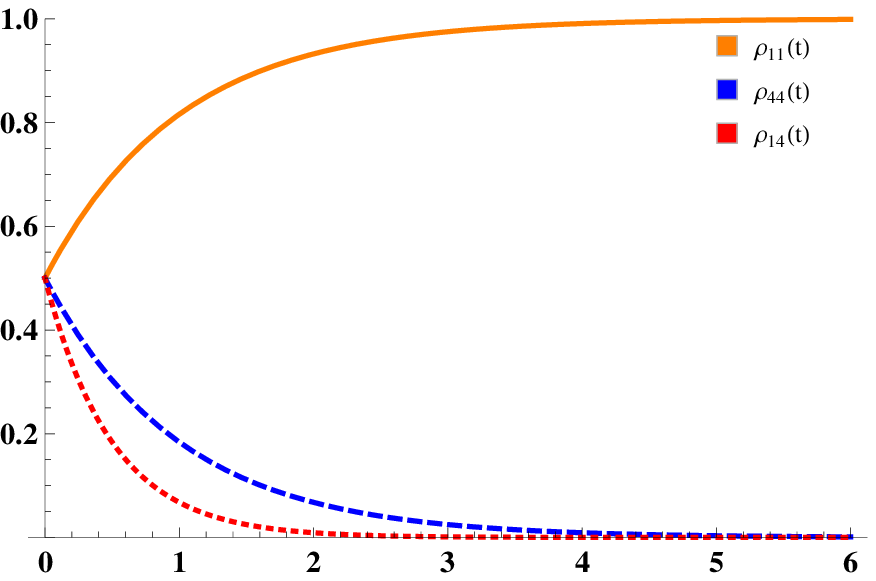}
	\put(2,-115){$\gamma_M t$}~~\quad
	\end{subfigure}
	\caption{The entanglement degrees of $\rho_{1}(t)$ vs $\gamma_M t$ in the Markovian regime, where $m_1=0$. Fig. (b) describes the evolution of different density matrix populations.}
\end{figure}

Fig.(2a) displays the evolution of concurrence, logarithmic negativity and quantum discord in the Markovian case given in Eq.(\ref{degree of entanglement 1}) against $\gamma_M t$. In  this case, we suppose that $a=b=c=d=1/\sqrt{2}$ and $n_1=m_1$. Initially it is obvious that these measures reach their maximum which is normal since $n_1=m_1=0$, i.e, the initial $|\phi\rangle_{AB}^1(0)$ is exactly the standard EPR state, that is, the state is initially maximally entangled. Once the dynamics is switched on ($t>0$), the entanglement degrees decrease gradually to completely vanish for $t\to \infty$ which means that joint field-field state inside cavities becomes separable. Moreover, a perfect similarity between the concurrence and  logarithmic negativity is clearly appeared. However, it is clear that the logarithmic negativity exceeds the concurrence, where the corresponding sudden death phenomenon of both of them appears fast comparing to those displayed for quantum discord. Fig.(2b) shows the evolution of various density matrix elements, namely $\rho_{11}(t)$, $\rho_{14}(t)$ and $\rho_{44}(t)$.  It is clear that the populations $\rho_{14}(t)$ and $\rho_{44}(t)$ decrease monotonically when $t\to\infty$, while $\rho_{11}(t)$ increases gradually to take stable behaviour for large numbers of $\gamma_M t$. This means that the populations $\rho_{44}(t)$ and $\rho_{14}(t)$ start to manifest which restrains the increasing population $\rho_{11}(t)$.\\

\begin{figure}[h!]
	\centering
	
	\begin{subfigure}[t]{0.03\textwidth}
		\textbf{(a)}
	\end{subfigure}
	\begin{subfigure}[t]{0.4\textwidth}
		\includegraphics[width=\linewidth,valign=t]{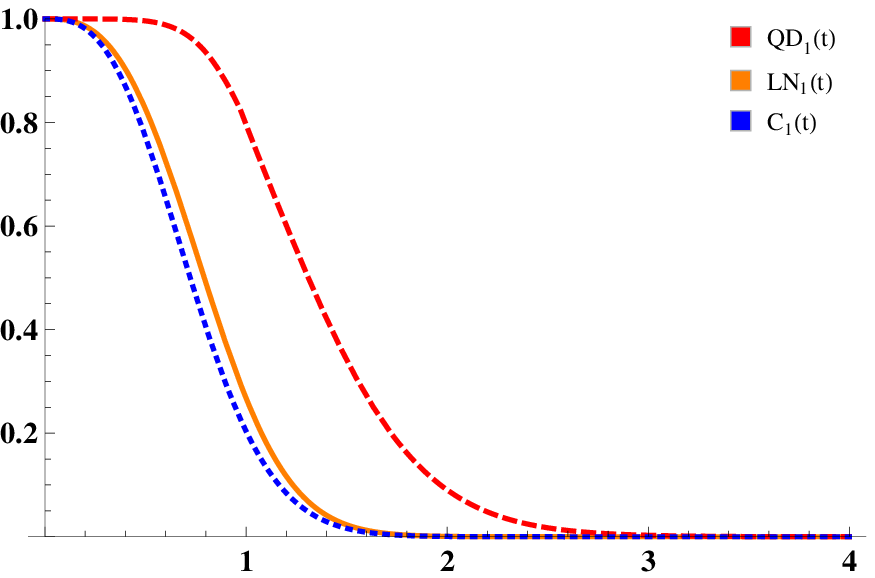}
		\put(-3,-120){$\omega_0 t$}~~\quad
	\end{subfigure}\hfill
	\begin{subfigure}[t]{0.03\textwidth}
		\textbf{(b)}  
	\end{subfigure}
	\begin{subfigure}[t]{0.4\textwidth}
		\includegraphics[width=\linewidth,valign=t]{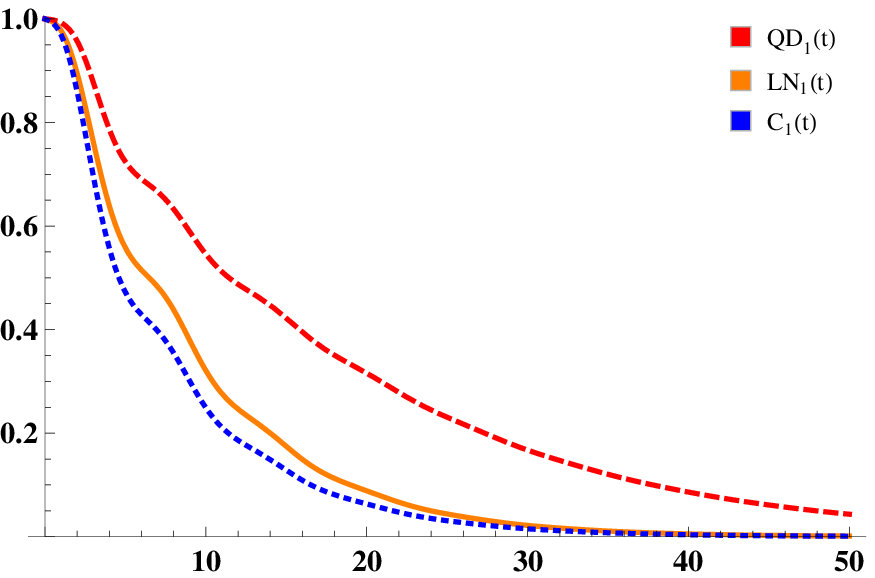}
		\put(-3,-120){$\omega_0 t$}~~\quad
	\end{subfigure}\hfill\\
	\begin{subfigure}[t]{0.03\textwidth}
	\textbf{(c)}
\end{subfigure}
\begin{subfigure}[t]{0.4\textwidth}
	\includegraphics[width=\linewidth,valign=t]{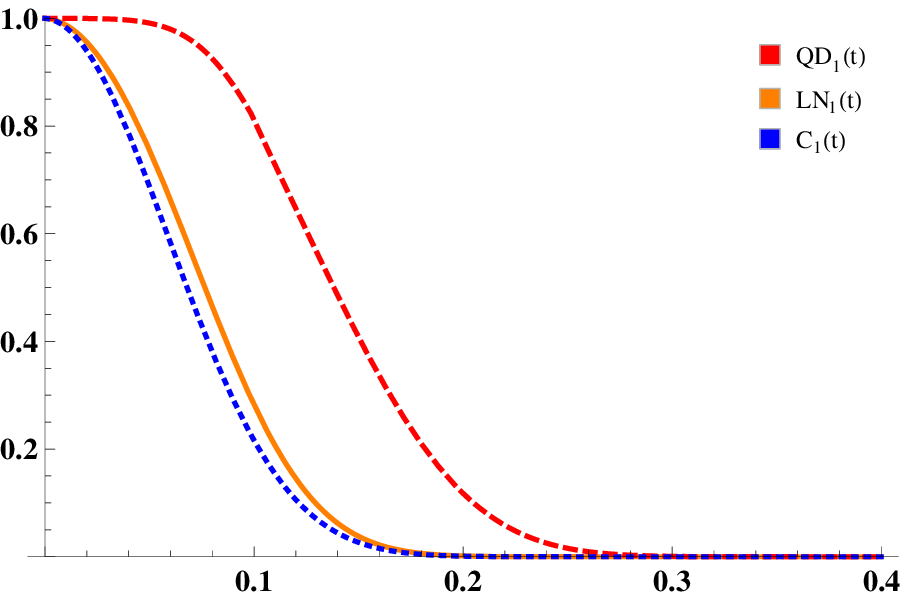}
		\put(5,-117){$\omega_0 t$}~~\quad
\end{subfigure}
\hfill
	\begin{subfigure}[t]{0.03\textwidth}
	\textbf{(d)}
\end{subfigure}
\begin{subfigure}[t]{0.4\textwidth}
	\includegraphics[width=\linewidth,valign=t]{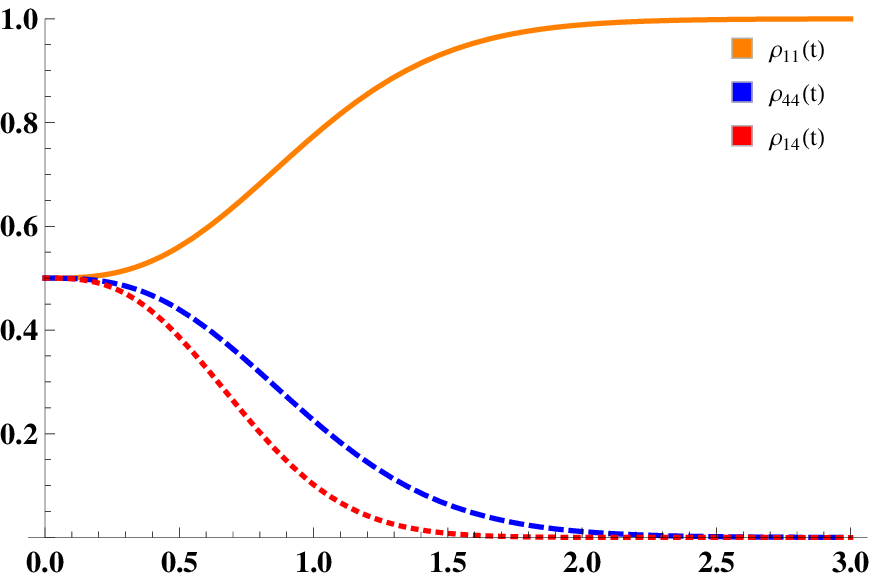}
		\put(5,-117){$\omega_0 t$}~~\quad
\end{subfigure}
	\caption{The entanglement degrees of $\rho_{1}(t)$ in the non-Markovian regime, where $m_1=0$ and (a) $ r = 1$, (b) $ r = 0,1$ and (c) $ r = 5$. Fig. (d) describes the evolution of different density matrix populations.}
\end{figure}

Similarly, in Fig. (3), we display the dynamics of the same quantities given in Eq.(\ref{degree of entanglement 1}) but when the non-Markovian regime is taken into account. Indeed, the solution of the non-Markovian master equation in Eq.(\ref{non-Markovian M.E}) allows to quantify the concurrence, logarithmic negativity and quantum discord for various numbers of $r$, namely $r=0.1,r=1$ and $r=5$, i.e, $\omega_c<\omega_0,  \omega_c=\omega_0 $ and $\omega_c>\omega_0$, respectively.   It is clear that for different initial settings of $r$ and $m_1$, the quantities $C_1(t)$, $LN_1(t)$ and $QD_1(t)$ take the maximum numbers at $\omega_0 t=0$, that the state $|\phi\rangle_{AB}^1$ is initially maximally entangled and fields inside the cavities A and B are strongly interacted with each other. For the further values of $\omega_0 t$, the witnesses of entanglement decrease gradually to vanish for $t\to \infty$. Again it is obvious that the logarithmic negativity fluctuates similarly as concurrence. However the plots show that the sudden death phenomenon depend on the variation of $r$, where in general one may conclude that for $ \omega_c=\omega_0$ the concurrence, logarithmic negativity and quantum discord vanishes quickly comparing to the case where $\omega_c<\omega_0$, while for $\omega_c>\omega_0$ they vanish fast  comparing  to $\omega_c>\omega_0$.\\

	\item \underline{Second case}: The second state consists of preparing the total initial field-field state  inside  the cavities $A$ and $B$ in a $NOON$ state of the form $|\phi\rangle _{AB} ^2(0)= b |n_1\rangle |m_1+1 \rangle+c |n_1+1\rangle |m_1\rangle $, that is, $a=d=0$ in Eq.(\ref{total initial state}). On the other hand, using the results in appendix $B$, one can  rewrite the state in Eq.(\ref{rho}) as follows,
\begin{equation}\label{rho2}
\rho^{2}(t)=\begin{pmatrix}
\rho_{11}(t)  & 0 & 0 & 0 \\
0 & \rho_{22}(t) & \rho_{23}(t) & 0 \\
0 & \rho_{23}^{*}(t) & \rho_{33}(t) & 0 \\
0 & 0 & 0 & 0
\end{pmatrix}.
\end{equation}
The concurrence, logarithmic negativity and quantum discord of the above joint field-field state (\ref{rho2}) are calculated respectively as bellow,
\begin{eqnarray}\label{enta2}
C_2(t)&=&\max(0,2[|\rho_{23}(t)|-\sqrt{\rho_{11}(t)}]),\nonumber\\
LN_2(t)&=&\max(0,\log_{2}[1-\rho_{11}(t)+\sqrt{\rho_{11}(t)^2+4|\rho_{23}(t)|^2}]),\nonumber\\
QD_{2}(t)&=&\min(Q_1^{2},Q_2^{2}),
\end{eqnarray}
where
\begin{eqnarray}
Q_1^{2}&=&H(\rho_{11}(t)+\rho_{33}(t))+ \sum_{j=1}^{4} \lambda_j \log_{2}\lambda_j +H(\frac{1}{2}(1+\sqrt{[1-2 \rho_{33}(t)]^2+4|\rho_{23}(t)|^2 })),\nonumber \\
Q_2^{2}&=& \sum_{i=1}^{4} \lambda_j \log_{2}\lambda_j+\sum_{i=1}^{4} \rho_{ii}.
\end{eqnarray}
 \end{enumerate}
Figs. (4a) and (4b) display the evolution of $C_2(t)$, $LN_2(t)$ and $QD_2(t)$ already calculated in Eq.(\ref{enta2}) in Markovian and non-Markovian cases, respectively. In both situations, it is clear that at $t=0$ and for different initial settings of $m_1$ and $r$, the joint field-field state, namely $\rho^{2}(t)$ is initially maximally entangled. For the further value of $t$, the entanglement measures decreases monotonically to reach their minimum values and vanish for large numbers of $\gamma_M t$ and $\omega_0 t$, i.e, the $\rho^{2}(t)$ is separable. Moreover, it is clear that the concurrence and logarithmic negativity fluctuate similarly between their minimum and maximum bounds, where the logarithmic negativity takes smaller numbers comparing to those obtained for concurrence. Obviously, a perfect similarity between quantum discord  and concurrence is appeared. Again, it is clear that  the sudden death phenomenon appears fast in the non-Markovian dynamic and it depends basically on the cavity  field parameters, namely  $\gamma$, $\omega_0$ and $m_1$.\\

 \begin{figure}[h!]
	\centering
	\begin{subfigure}[t]{0.03\textwidth}
		\textbf{(a)}
	\end{subfigure}
	\begin{subfigure}[t]{0.4\textwidth}
		\includegraphics[width=\linewidth,valign=t]{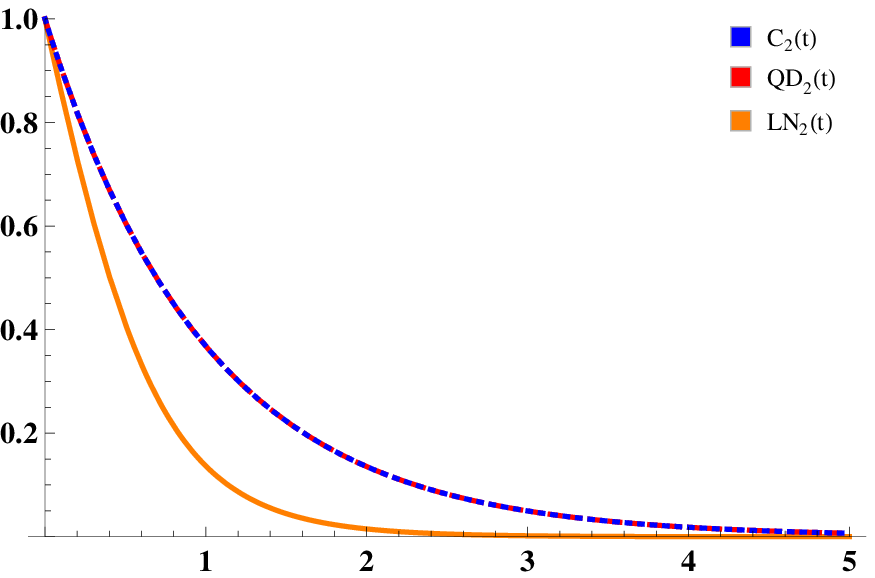}
		\put(-3,-120){$\gamma_M t$}~~\quad
	\end{subfigure}\hfill
	\begin{subfigure}[t]{0.03\textwidth}
		\textbf{(b)}  
	\end{subfigure}
	\begin{subfigure}[t]{0.4\textwidth}
		\includegraphics[width=\linewidth,valign=t]{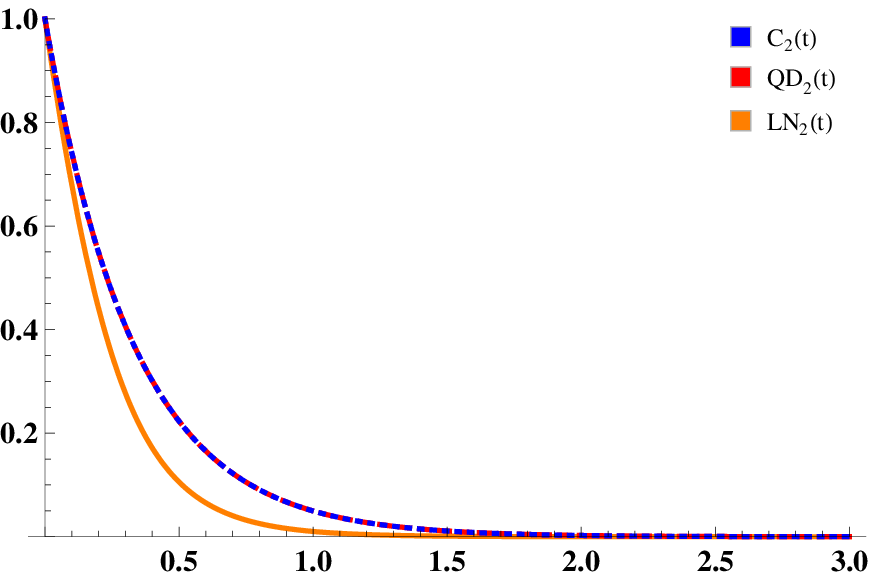}
		\put(-3,-120){$\gamma_M t$}~~\quad
	\end{subfigure}\hfill\\
\begin{subfigure}[t]{0.03\textwidth}
	\textbf{(c)}
\end{subfigure}
\begin{subfigure}[t]{0.4\textwidth}
	\includegraphics[width=\linewidth,valign=t]{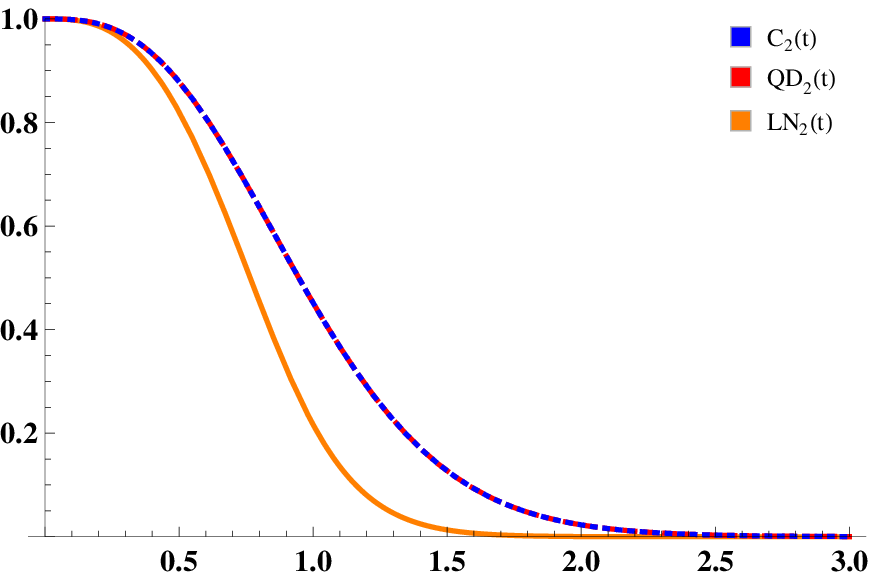}
	\put(-3,-120){$\omega_0 t$}~~\quad
\end{subfigure}\hfill
\begin{subfigure}[t]{0.03\textwidth}
	\textbf{(d)}  
\end{subfigure}
\begin{subfigure}[t]{0.4\textwidth}
	\includegraphics[width=\linewidth,valign=t]{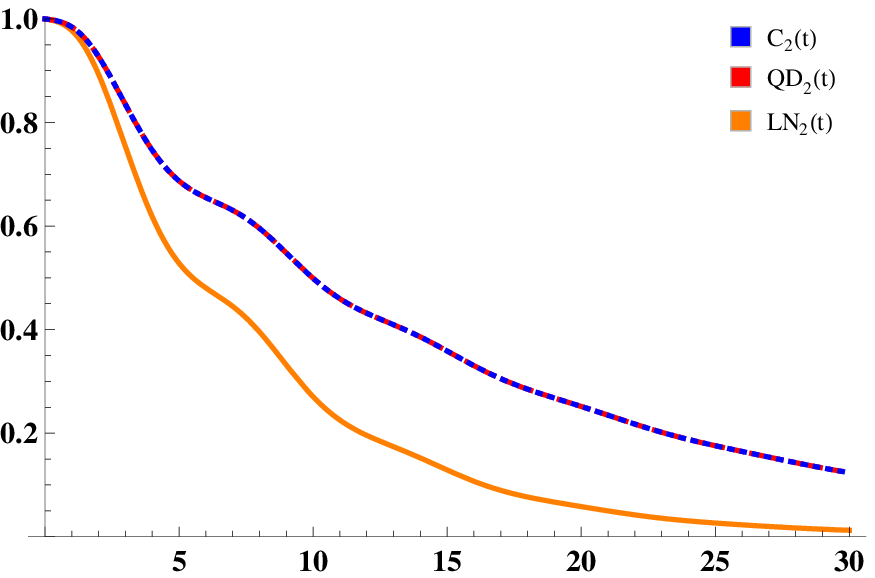}
	\put(-3,-120){$\omega_0 t$}~~\quad
\end{subfigure}
	\caption{The entanglement degrees of $\rho_{2}(t)$ in the Markovian regime, where (a) $ m_1 = 0$, (b) $ m_1 = 1$. Fig. (c) and display the same quantities but in the non-Markovian regime, where $ m_1 = 0$ and (c) $r = 1$, (d) $ r = 0.1$, respectively.}
\end{figure}
From Figs. (2-4), the general analysis show that our proposed model allows to quantify directly the separability between the component of a bipartite field-field state using EPR and NOON states as an initial states. By controlling the cavity field parameters, it is found that initially the total field-field state is maximally entangled which means that its component are initially strongly coupled. Once the dynamics is switched on the bipartite state becomes separable for large interval of time. However, it is shown that the sudden death time phenomenon in both dynamics, namely the Markovian and non-Markovian regimes depends in general of the cavity field parameters. From another perspective, it would be motivating to shed light on a further tool extensively used to quantify the rate of entanglement inherent in quantum state which is negativity volume based on the Wigner function calculated in Eq. (\ref{volu}). Indeed, Fig. (5) displays the evolution of the Wigner function and  negativity volume, namely $W$ and $V$, respectively of the density matrix given in Eq.(\ref{rho}), where the Markovian and non-Markovian regimes are considered. It is clear that for various numbers of $m_1$, the Wigner function decreases monotonically  when $t$ increases and never vanishes, however the negativity volume increases  as $\gamma_M t\to\infty$. Again these indicators prove that initially,the joint field-field state is maximally entangled which means that its subsystems are strongly correlated, while for the various values of t the state becomes separable. Hence, since the total field-field state given in Eq. (\ref{rho}) generate a nonzero negativity volume, then it reflects the  non-classicality phenomenon which prove again that the field inside the cavities $A$ and $B$ are entangled.\\ 

\begin{figure}[h!]
	\centering
	\begin{subfigure}[t]{0.03\textwidth}
		\textbf{(a)}
	\end{subfigure}
	\begin{subfigure}[t]{0.4\textwidth}
		\includegraphics[width=\linewidth,valign=t]{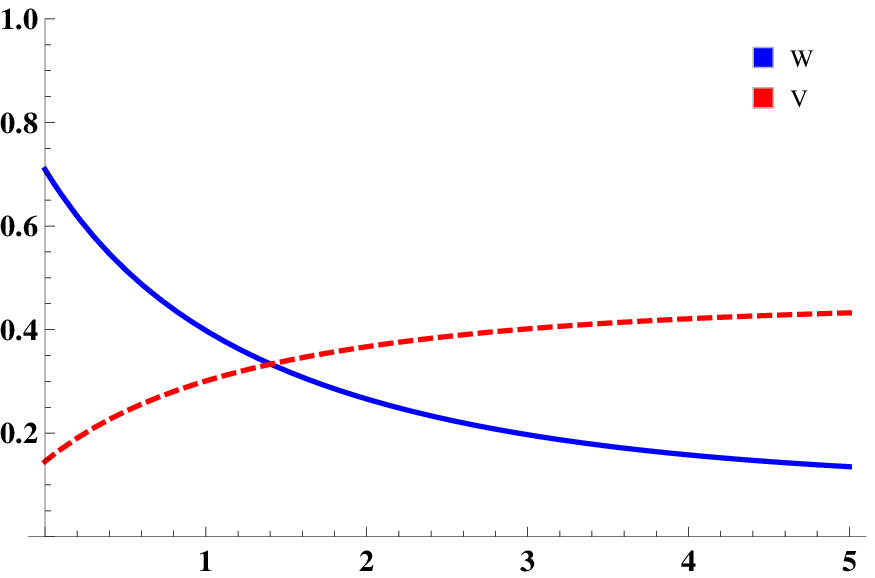}
		\put(-3,-120){$\gamma_M t$}~~\quad
	\end{subfigure}\hfill
	\begin{subfigure}[t]{0.03\textwidth}
		\textbf{(b)}
	\end{subfigure}
	\begin{subfigure}[t]{0.4\textwidth}
		\includegraphics[width=\linewidth,valign=t]{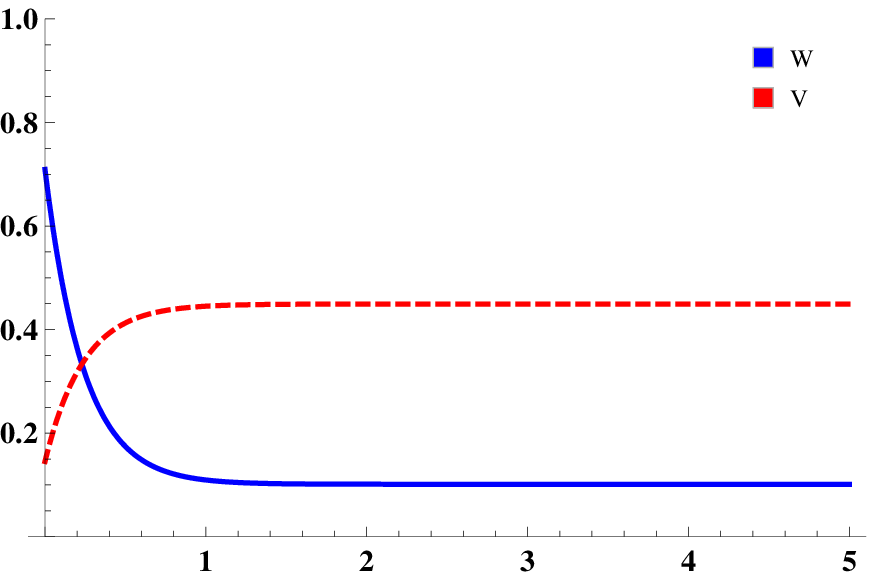}
		\put(-3,-120){$\gamma_M t$}~~\quad
	\end{subfigure}	\hfill\\
	\begin{subfigure}[t]{0.03\textwidth}
		\textbf{(c)}
	\end{subfigure}
	\begin{subfigure}[t]{0.4\textwidth}
		\includegraphics[width=\linewidth,valign=t]{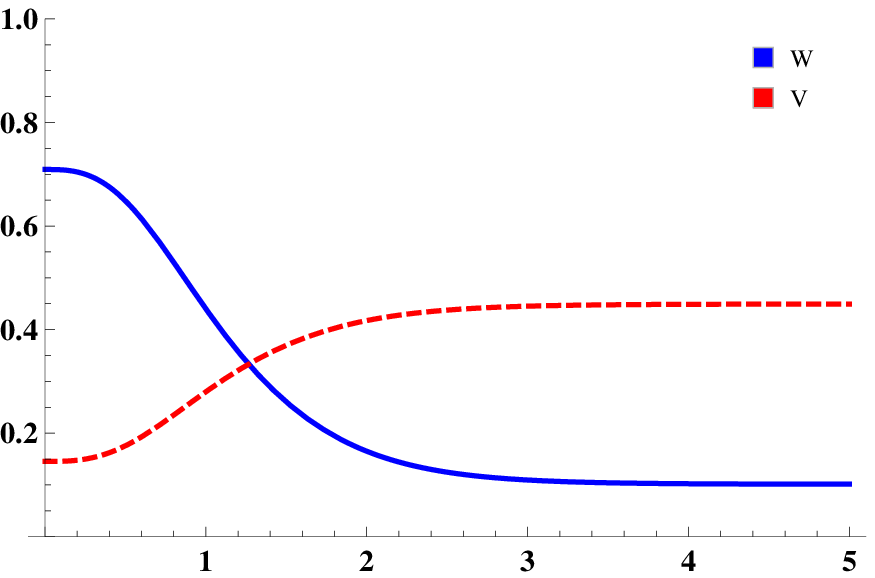}
		\put(-3,-120){$\omega_0 t$}~~\quad
	\end{subfigure}\hfill
	\begin{subfigure}[t]{0.03\textwidth}
		\textbf{(d)}
	\end{subfigure}
	\begin{subfigure}[t]{0.4\textwidth}
		\includegraphics[width=\linewidth,valign=t]{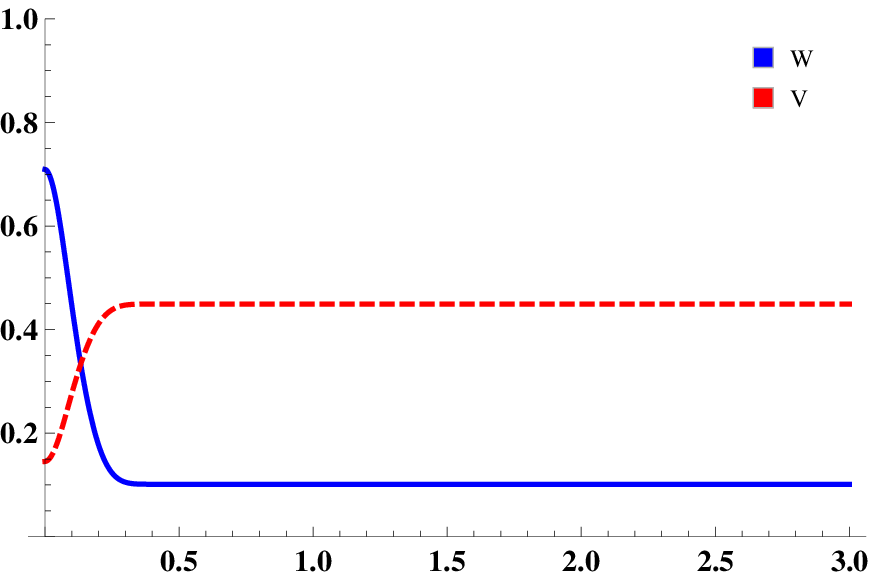}
		\put(-3,-120){$\omega_0 t$}~~\quad
	\end{subfigure}
	\caption{Dynamics of the Wigner function and negativity volume in the Markovian regime, where (a) $m_1=0$, (b) $m_1=2$. Fig. (c) and (d) display the same quantities but in  the non-Markovian regime, where $r = 1$ and (c) $m_1=0$, (d) $m_1=2$.}
\end{figure}

From the above analysis since the states $\rho^{1}(t)$ and $\rho^2(t)$ are entangled then they can be considered as a good partial-entangled quantum channels in the process of quantum teleportation. Indeed in the next section we shall propose two schemes of quantum teleportation to teleport a arbitrary bipartite quantum state.

\section{Quantum teleportation protocol}
The main subject of this protocol is to send an arbitrary quantum state from a sender to a distant receiver, named Alice and Bob, respectively. Moreover, these partners  are usually connected with each other by a partial (maximally) entangled channel. In our case we assume that Alice and Bob are connected via the entangled joint field state in Eq.(\ref{rho1}) and  Eq.(\ref{rho2}). However, suppose that the state to be teleported, namely $\rho_{un}=|\psi_{un}\rangle \langle \psi_{un} |$ takes the following compact form, 
  \begin{equation}\label{unknown state}
  |\psi_{un}\rangle=\frac{1-2p}{2}|00\rangle \langle00|+\frac{1+2p}{2}|11\rangle \langle11|+\frac{q}{2}(|11\rangle \langle 00|+|00\rangle \langle 11|),
  \end{equation}
where $0\le p \le 1$ and $q=\sqrt{1-p^2}>0$. First of all, Alice performs a generalized measurements on her own state. The second step consists of communicating her results via a classical channel with Bob. At the end of this process and after performing a unitary transformation, the final teleported state can written as bellow \cite{35}

\begin{equation}
\rho_{out}=\sum_{\alpha \beta} P_{\alpha \beta} (\sigma_\alpha \otimes \sigma_\beta)\rho_{un} (\sigma_\beta\otimes \sigma_\alpha ),
\end{equation}
where $P_{\alpha \beta_{1(2)}}= Tr[E^\alpha\rho_{1(2)}(t)] Tr[E^\beta\rho_{1(2)}(t)]$, and $\sigma_{\alpha \beta}(\alpha, \beta=0,x,y,z)$ denote the three components of Pauli matrices, while $\sigma_0$ denotes the identity operator. Moreover, the operators $E^{\alpha}$ are connected to the Bell states as follows,
\begin{equation}
E ^{0/z}=|\psi^{-/+}\rangle \langle \psi^{-/+}|\quad,\quad
E ^{x/y}=|\phi^{-/+}\rangle \langle \phi^{-/+}|
\end{equation}

\begin{equation}
 |\psi^{\pm}\rangle=\frac{|01\rangle+|10\rangle}{\sqrt{2}}\quad,\quad
| \phi^{\pm}\rangle=\frac{|00\rangle+|11\rangle}{\sqrt{2}}.
\end{equation}
Moreover, the quality of Bob's state is evaluated using the so-called fidelity which is defined as $F=Tr[\rho_{un} \rho_{out}]$. Indeed, the fidelity reflects the credibility between the input and the output states, i.e, it measures how close the final state to the initial state. However, it is equal to one if the teleportation process is perfect (maximally entangled channel). Otherwise, if the channel is partially entangled, then $0 \le F \le 1$.
\begin{itemize}
	\item \textbf{Case of the partial entangled channel $\rho^{1}(t)$}
	
The aim of Alice is to sends the unknown state in Eq.(\ref{unknown state}) to Bob, where the entangled state defined in Eq.(\ref{rho1}) is taken to be the quantum channel between them. At the end of the process Bob has the output final state of the form

	\begin{equation}\label{teleported state 1}
	\rho_{out}^1(t)=\begin{pmatrix}
	k_1(t)  & 0 & 0 & 	k_2(t)  \\
	0 & 0 & 	k_3(t)  & 0 \\
	0 & 	k_3(t)  & 0 & 0 \\
		k_2(t)  & 0 & 0 & 	k_1(t)
	\end{pmatrix} ,
	\end{equation}
	where,
	\begin{eqnarray}
		k_1(t) &=&	\frac{1-2p}{2}(\rho_{22}(t)+\rho_{33}(t))^2+\frac{1+2p}{2}(\rho_{11}(t)+\rho_{44}(t))^2,\quad k_{2}(t)=2q\rho_{14}(t)^2,\nonumber\\
		k_{3}(t)&=&(\rho_{11}(t)+\rho_{44}(t))(\rho_{22}(t)+\rho_{33}(t)).
	\end{eqnarray}
The elements $\rho_{11}(t),\rho_{22}(t),\rho_{33}(t),\rho_{44}(t)$ and $\rho_{14}(t)$ are given in Appendix $B$. The fidelity $F_1$ and the teleported entanglement degrees are calculated, respectively as follows,

\begin{eqnarray}
F_1&=&k_1(t)+qk_2(t),\nonumber\\
C_{out}^1(t)&=&\max \{0,2(k_3(t)-k_1(t)),2k_2(t)\},\nonumber\\
LN_{out}^1(t)&=&\max  \{0,\log_{2}(1+2[k_2(t)+k_3(t)-k_1(t)])\},\nonumber\\
QD_{out}^1(t)&=&\min(Q_{out}^{1},Q_{out}^{1^{'}}),
\end{eqnarray}

where
\begin{eqnarray}
Q_{out}^{1}&=&H(k_{1}(t))+ \sum_{j=1}^{4} \lambda_j \log_{2}\lambda_j+H(\frac{1}{2}(1+\sqrt{[1-2 k_{1}(t)]^2+4[k_2(t)+k_3(t)]^2 })),\nonumber\\
Q_{out}^{1^{'}}&=& \sum_{i=1}^{4} \lambda_j \log_{2}\lambda_j+2 k_1(t),
\end{eqnarray}
where $\lambda_j$ denote the eigenvalues of $\rho_{out}^1(t)$.\\

\begin{figure}[h!]
	\centering
	
	\begin{subfigure}[t]{0.03\textwidth}
		\textbf{(a)}
	\end{subfigure}
	\begin{subfigure}[t]{0.4\textwidth}
		\includegraphics[width=\linewidth,valign=t]{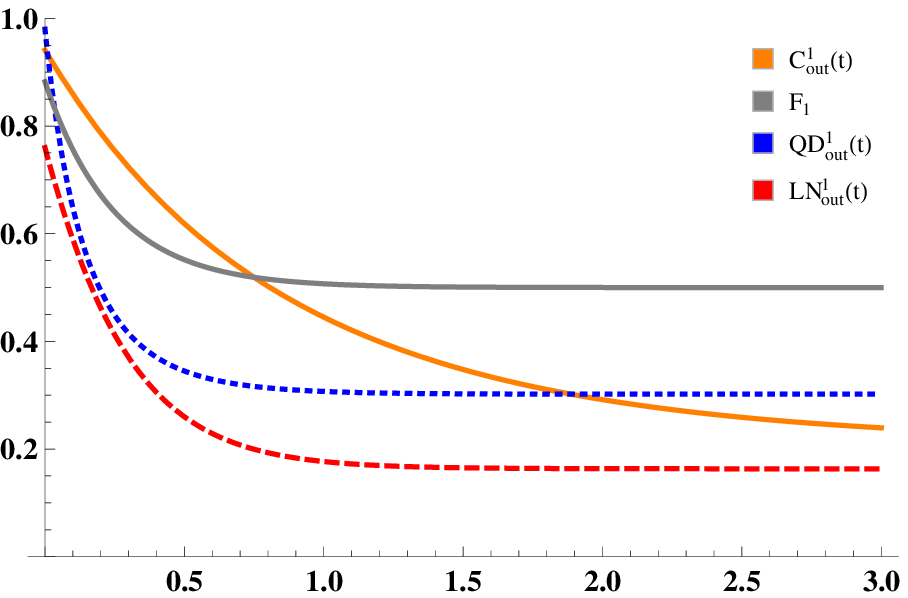}
		\put(3,-115){$\gamma_M t$}~~\quad
	\end{subfigure}\hfill
	\begin{subfigure}[t]{0.03\textwidth}
		\textbf{(b)}  
	\end{subfigure}
	\begin{subfigure}[t]{0.4\textwidth}
		\includegraphics[width=\linewidth,valign=t]{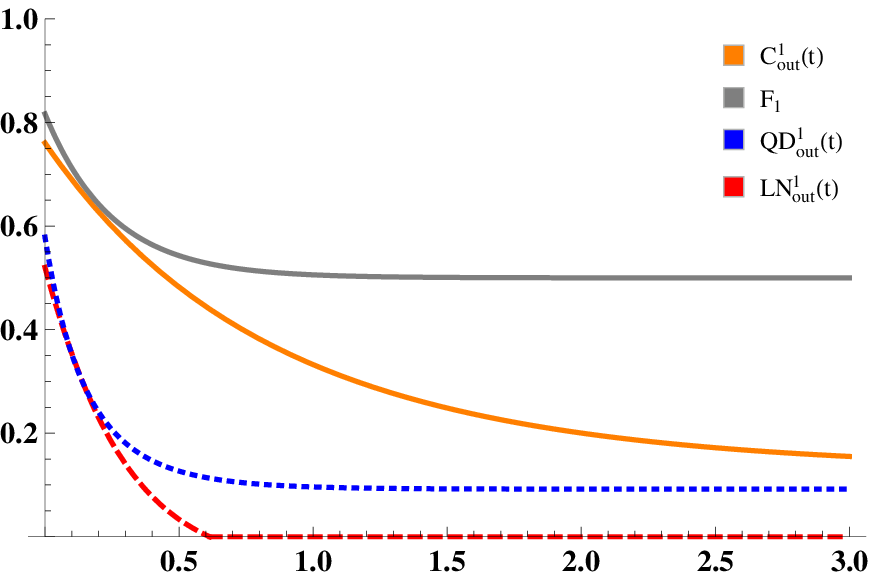}
	    \put(-2,-119){$\gamma_M t$}~~\quad
	\end{subfigure}
	\caption{The entanglement degree of $\rho_{out}^1(t)$ in the Markovian regime, where $a=b=c=d=\frac{1}{\sqrt{2}}$, $m_1=0$ and (a)   $p=0.99$, (b) $q=0.97$.}
\end{figure}
	Figs. (6) and (7) show the evolution of concurrence, logarithmic negativity, quantum discord and fidelity of the teleported state in Markovian and non-Markovian cases, respectively for various numbers of $p$, $q$ and $m_1$. Initially, the behaviours show that the teleported entanglement degrees and fidelity, namely $C_{out}^1$, $N_{out}^1$, $QD_{out}^1$ and $F_1$, respectively  reach the maximum bounds which is almost $1$, i.e, the input and output states are initially identical, where the transmission credibility of the unknown quantum state measured in terms of $F_1$ is then perfect. For the further values of t the teleported  entanglement degrees as well as fidelity  decrease for large values and smaller numbers of  $m_1$  and $q$ decrease. However, it is obvious that the maximum value of $F_1$  exceeds $\frac{2}{3}$ and it never vanishes which means that we arrived to exceed the maximum classical information. Therefore, one may considered $\rho_{1}(t)$ as a good quantum channel under this process.
 \begin{figure}[h!]
 	\centering
 	
 	\begin{subfigure}[t]{0.03\textwidth}
 		\textbf{(a)}
 	\end{subfigure}
 	\begin{subfigure}[t]{0.4\textwidth}
 		\includegraphics[width=\linewidth,valign=t]{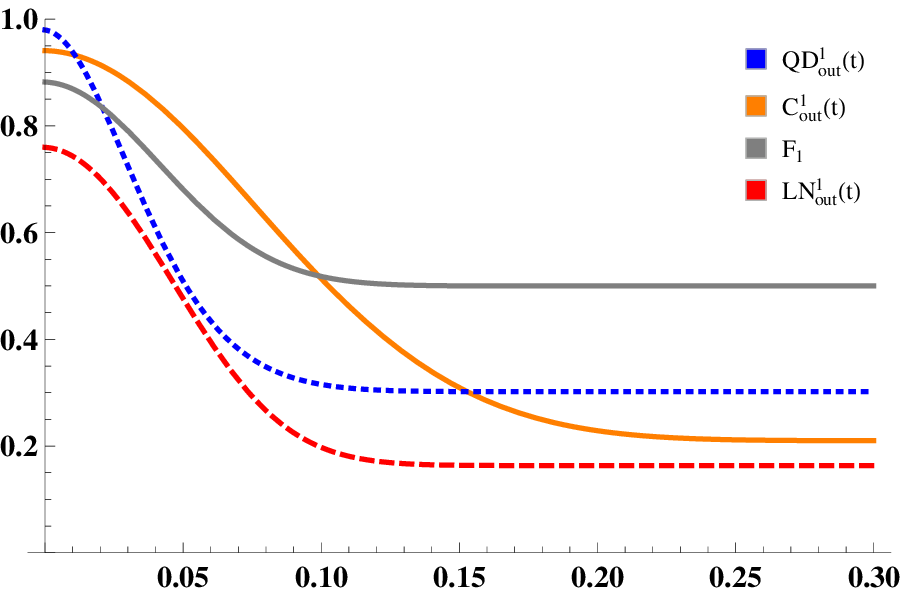}
 			\put(3,-115){$\omega_0 t$}~~\quad
 	\end{subfigure}\hfill
 	\begin{subfigure}[t]{0.03\textwidth}
 		\textbf{(b)}  
 	\end{subfigure}
 	\begin{subfigure}[t]{0.4\textwidth}
 		\includegraphics[width=\linewidth,valign=t]{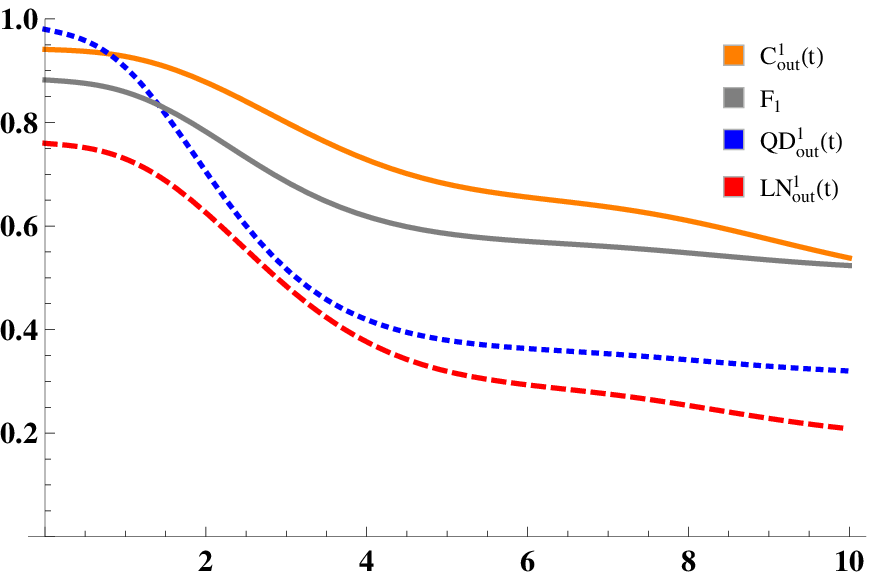}
 				\put(3,-115){$\omega_0 t$}~~\quad
 	\end{subfigure}\hfill\\
 	\begin{subfigure}[t]{0.03\textwidth}
 		\textbf{(c)}
 	\end{subfigure}
 	\begin{subfigure}[t]{0.4\textwidth}
 		\includegraphics[width=\linewidth,valign=t]{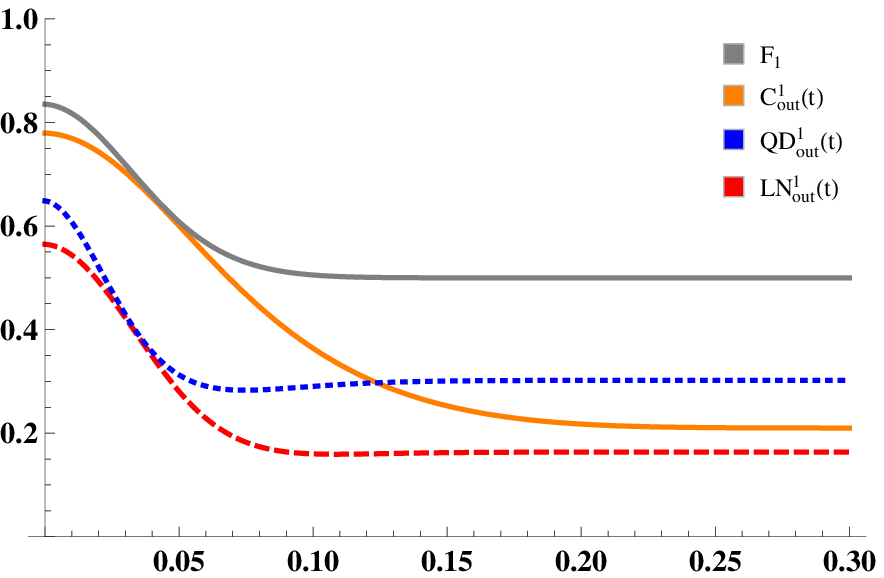}
 				\put(-2,-119){$\omega_0 t$}~~\quad
 	\end{subfigure}
 	\caption{The entanglement degrees of $\rho_{out}^1(t)$ in the non-Markovian regime, where $a=b=c=d=\frac{1}{\sqrt{2}}$, $m_1=0$, $p=0.99$, $q=0.97$ and (a) $r=1$, (b) $r=0.1$, (c) $r=5$.}
 \end{figure}
\item  \textbf{Case of partial entangled channel with $\rho_{2}(t)$}

In this case we suppose that  Alice sends the same unknown state (\ref{unknown state}) to Bob. We assume that the partners, namely Alice and Bob are connected using the quantum channel in Eq.(\ref{rho2}). At the end of the process Bob has the output state as bellows,
\begin{equation}
\rho_{out}^2(t)=\begin{pmatrix}
\alpha_1(t)  & 0 & 0 & 	\alpha_2(t)  \\
0 & 0 & 	\alpha_3(t)  & 0 \\
0 & 	\alpha_3(t)  & 0 & 0 \\
\alpha_2(t)  & 0 & 0 & 	\alpha_1(t)
\end{pmatrix} ,
\end{equation}
where
\begin{eqnarray}
\alpha_1(t) &=&	(\rho_{22}(t)+\rho_{33}(t))^2\frac{1-2p}{2}+\rho_{11}(t)^2\frac{1+2p}{2},\quad
\alpha_{2}(t)=2q\rho_{23}(t)^2,\quad
\alpha_{3}(t)=\rho_{11}(t)(\rho_{22}(t)+\rho_{33}(t)).
\end{eqnarray}
The fidelity and the teleported entanglement degrees are calculated, respectively as

\begin{eqnarray}
F_2&=&\alpha_1(t)+q\alpha_2(t),\nonumber\\
C_{out}^1(t)&=&\max \{0,2(\alpha_3(t)-k_1(t)),2\alpha_2(t)\},\nonumber\\
LN_{out}^1(t)&=&\max  \{0,\log_{2}(1+2[\alpha_2(t)+\alpha_3(t)-\alpha_1(t)]) \},\nonumber\\
QD_{out}^1(t)&=&\min(Q_{out}^{2},Q_{out}^{2^{'}}),
\end{eqnarray}
where
\begin{eqnarray}
Q_{out}^{2}&=&H(\alpha_{1}(t))+ \sum_{j=1}^{4} \lambda_j \log_{2}\lambda_j +H(\frac{1}{2}(1+\sqrt{[1-2 \alpha_{1}(t)]^2+4[\alpha_2(t)+\alpha_3(t)]^2 })),\nonumber\\
Q_{out}^{2^{'}}&=& \sum_{l=1}^{4} \lambda_l \log_{2}\lambda_l+2 \alpha_{1}(t),
\end{eqnarray}
where $\lambda_l$ are the eigenvalues of $\rho_{out}^2(t)$.

\begin{figure}[h!]
	\centering
	
	\begin{subfigure}[t]{0.03\textwidth}
		\textbf{(a)}
	\end{subfigure}
	\begin{subfigure}[t]{0.4\textwidth}
		\includegraphics[width=\linewidth,valign=t]{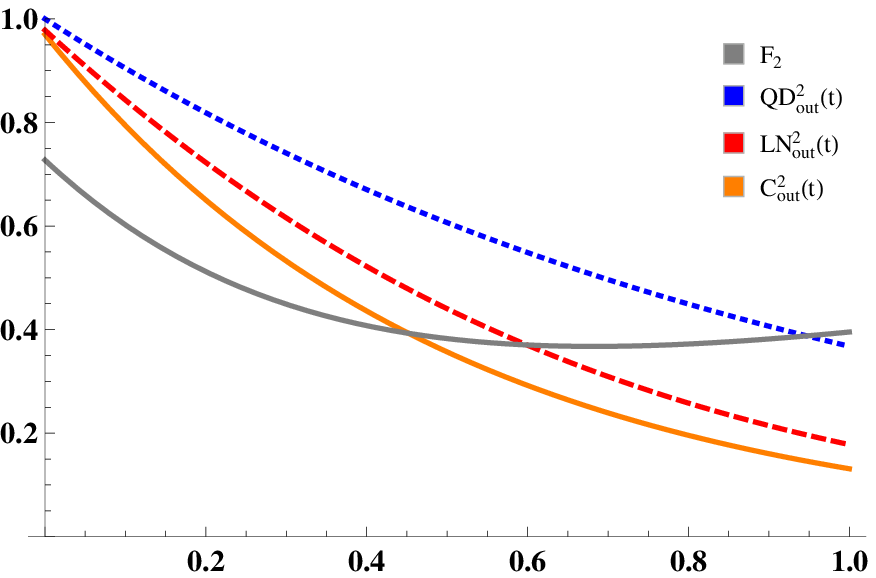}
			\put(-3,-119){$\gamma_M t$}~~\quad
	\end{subfigure}\hfill
	\begin{subfigure}[t]{0.03\textwidth}
		\textbf{(b)}  
	\end{subfigure}
	\begin{subfigure}[t]{0.4\textwidth}
		\includegraphics[width=\linewidth,valign=t]{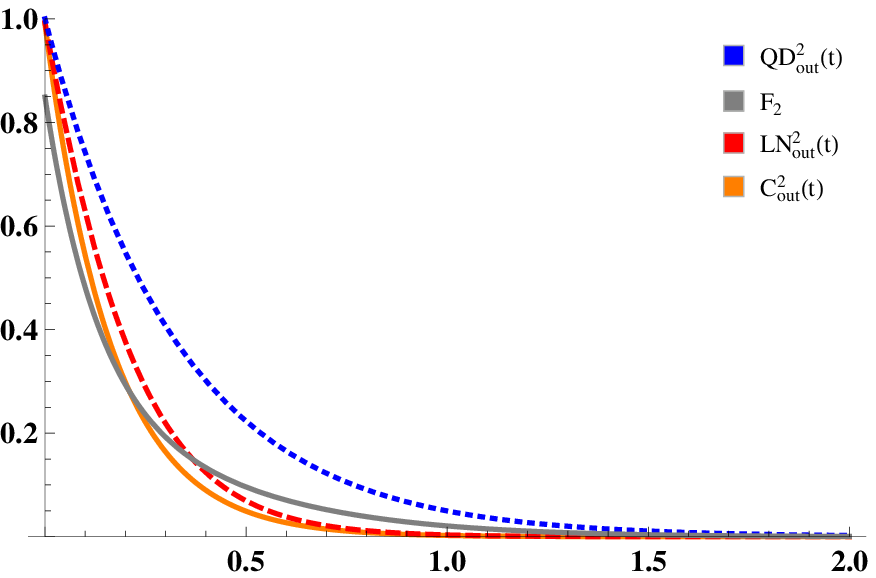}
			\put(-3,-119){$\gamma_M t$}~~\quad
	\end{subfigure}
	\caption{The entanglement degree of $\rho_{out}^2(t)$ in the Markovian regime, where $a=b=c=d=\frac{1}{\sqrt{2}}$ and  (a) $q=0.97$, $m_1=0$ ,(b) $q=0.99$,  $m_1=1$.}
\end{figure}
In  Figs. (8) and (9) we investigate the evolution of the entanglement degrees and fidelity of the second teleported state, namely $\rho_{out}^2(t)$ in the Markovian and non-Markovian regimes, respectively. Both behaviours show a good similarity between concurrence and logarithmic negativity since they have the same dimensionality. Moreover, it is clear that we arrived to exceed the maximum of classical information , i.e, $F_2 \ge \frac{2}{3} $. However, if $m_1=0$, that is, the second initial state is prepared to be $|\psi_{AB}\rangle (0) =b|0\rangle |1\rangle +c |1 \rangle |0\rangle $ which is a $NOON$ state, then one can obtain the maximum value of concurrence, logarithmic negativity, quantum discord and also robust fidelity. Obviously, the sudden death time phenomenon appears fast in non-Markovian dynamics when $m_1$ and $r$  take large numbers. However, for the small value of $r$ ($r=0.1$) this  phenomenon appears when $t\to \infty$. Moreover, from both figures one can conclude that the maximum  bound of $F_2$ exceeds $\frac{2}{3}$ which means that the partial-entangled state $\rho^{2}(t)$ can be also used as a good quantum channel in quantum teleportation protocol.\\

\begin{figure}[h!]
	\begin{subfigure}[t]{0.03\textwidth}
		\textbf{(a)}
	\end{subfigure}
	\begin{subfigure}[t]{0.35\textwidth}
		\includegraphics[width=\linewidth,valign=t]{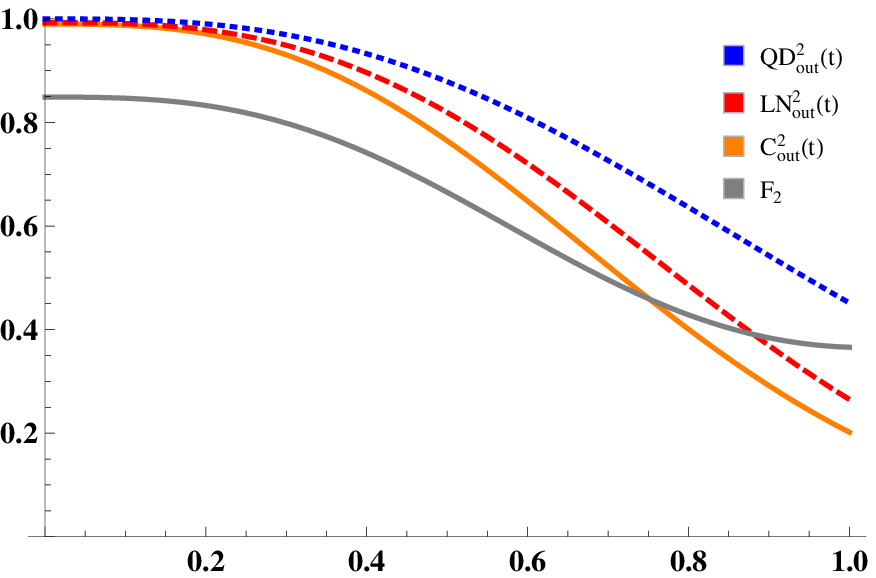}
				\put(-3,-119){$\omega_0 t$}~~\quad
	\end{subfigure}
	\begin{subfigure}[t]{0.03\textwidth}
		\textbf{(b)}  
	\end{subfigure}
	\begin{subfigure}[t]{0.33\textwidth}
		\includegraphics[width=\linewidth,valign=t]{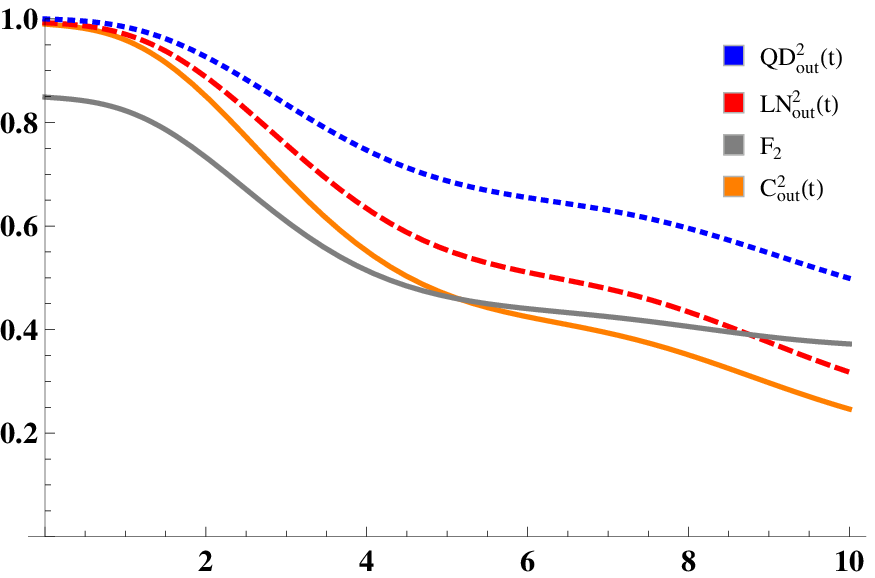}
			\put(-3,-119){$\omega_0 t$}
			\includegraphics[width=\linewidth,valign=t]{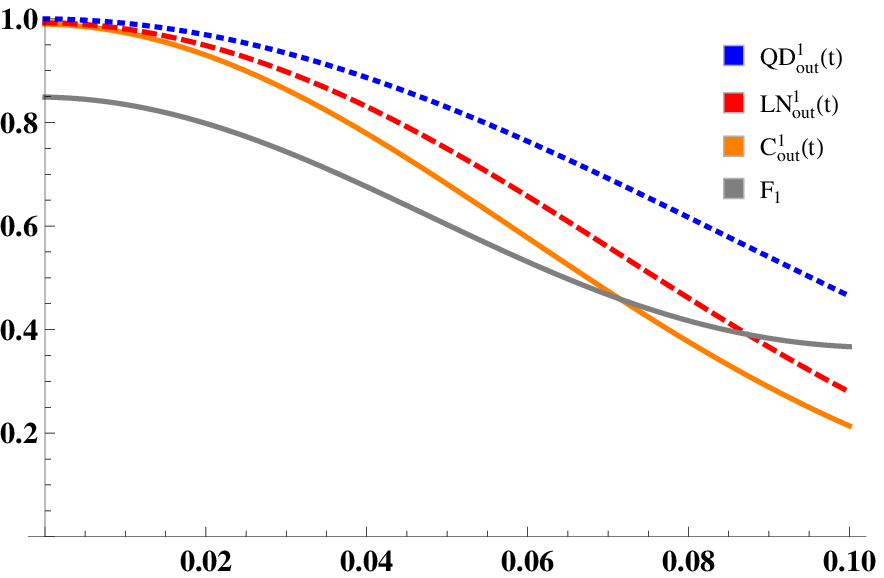}
			\put(-3,-119){$\omega_0 t$}~~\quad
	\end{subfigure}
	\caption{The entanglement degree of $\rho_{out}^2(t)$ in the non-Markovian regime, where $a=b=c=d=\frac{1}{\sqrt{2}}$, $m_1=0$, $q=0.99$ and (a) $r=1$, (b) $r=0.1$, (c) $r=5$.}
\end{figure}
\end{itemize}

It is worth emphasizing that in the above contribution we have providing an exact solution of Lindblad master equation of a joint field-field system inside two serial cavities A and B in Markovian and non-Markovian regimes using a novel general method. Indeed, S. Bougouffa in his paper \cite{29} provides a solution of the master equation (\ref{Lindblad M.E}) using the basis $\{|00\rangle, |01\rangle, |10\rangle, |11\rangle\}$. Building on that and by using the Fock basis arrived from the coherent states described the field inside the cavities A and B we have solved the Lindblad master equation but by using the basis $\{ |n_1,m_1\rangle, |n_1,m_1+1 \rangle, |n_1+1,m_1\rangle, |n_1+1,m_1+1\rangle \}$. In particular if we put $n_1=m_1=0$, then our results gives rise exactly to the same results in \cite{29}. Moreover, in this work we mainly focus our attention to solve the Lindblad master equation in Markovian and non-Markovian regimes. To benefit from our results, we have studied the entanglement rate inherent in the reduced field-field density matrix using various witnesses f entanglement. Finally as an application we proposed two schemes of quantum teleportation in order to study the credibility of transmitting an arbitrary bipartite quantum state between two partners Alice and Bob.
\section{Conclusion}

In this contribution, we have studied a physical  model based on the interaction between the field generated from  an excited atom passes through two identical cavities $A$ and $B$ and the field inside cavities. We have solved the Lindblad master equation describing the reduced density matrix of the joint field-field inside the cavities $A$ and $B$ in Markovian and non-Markovian dynamics using a bipartite coherent state to describe the field inside each cavity. By controlling the cavity field parameters and for two different kinds of entangled states, namely $EPR$ and $NOON$ states we have investigated the evolution of entanglement degrees of the resulting reduced density matrices of the field inside cavities. Moreover we have investigated the evolution of the Wigner function as well as the negativity volume of the joint field-field density matrix in order to display its non-classicality.  A comparative study between these witness shows that the state is entangled in both cases, namely Markovian and non-Markovian regimes. Moreover, we find that the sudden death time phenomenon appears fast in the non-Markovian case. Finally, using the obtained entangled field-field states as quantum channels we have proposed two schemes of quantum teleportation protocol. It is shown that the teleported entanglement degrees as well as fidelity depend in general of the cavity field parameters. These results can be used theoretically and experimentally in many tasks in quantum information theory, quantum metrology, quantum sensing, etc. Our future perspective is going to study other types of reservoirs which may allow to improve the cavity-QED technologies. Moreover it would be interesting to move to the high dimension , namely $2\times 3$ and $3\times 3$ systems to study entanglement, decoherence, teleportation protocol and other tasks in quantum information theory.

\section*{Appendix A. }

The master equations (\ref{Markovian M.E}) and (\ref{non-Markovian M.E}), are equivalent to a system of coupled differential equation which are calculated as bellow,
\begin{eqnarray}\label{y}
\frac{\rho_{11}}{dt}&=&[- \theta (2 \bar{n}(n_1+m_1+1)+(n_1+m_1)] \rho_{11}(t)\nonumber\\
&+&\theta(\bar{n}+1)[(n_1+1) \rho_{33}(t)+(m_1+1) \rho_{22}(t)], \nonumber\\
\frac{\rho_{12}}{dt}&=&-\frac{\theta}{2} (\bar{n}+1) [(2n_1+2m_1+1)\rho_{12}(t)-2(n_1+1)\rho_{34}(t)]\nonumber\\
&-&\frac{\theta}{2}\bar{n}(2n_1+m_1+3)\rho_{12}(t),\nonumber\\
\frac{\rho_{13}}{dt}&=&-\frac{\theta}{2} (\bar{n}+1) [(2n_1+2m_1+1)\rho_{13}(t)-2(n_1+1)\rho_{24}(t)]\nonumber\\
&-&\frac{\theta}{2}(2n_1+m_1+3)\rho_{13}(t),\nonumber\\
\frac{\rho_{14}}{dt}&=&-\theta(n_1+1)(n_1+m_1+1)\rho_{14}(t)\nonumber\\
&-&\frac{\theta}{2}\bar{n}(n_1+m_1+2)\rho_{14}(t),\nonumber\\
\frac{\rho_{21}}{dt}&=&-\frac{\theta}{2} (\bar{n}+1) [(2n_1+2m_1+1)\rho_{21}(t)-2(n_1+1)\rho_{43}(t)]\nonumber\\
&-&\frac{\theta}{2} \bar{n}(2n_1+m_1+3)\rho_{21}(t),\nonumber\\
\frac{\rho_{22}}{dt}&=&-\theta(\bar{n}+1)[(n_1+m_1+1)\rho_{22}(t)-(n_1+1)\rho_{44}(t)]\nonumber\\
&-&\theta\bar{n}[(n_1+1)\rho_{22}(t)-(m_1+1)\rho_{11}(t)],\nonumber\\
\frac{\rho_{23}}{dt}&=&-\theta (\bar{n}+1)(n_1+m_1+1)\rho_{23}(t)\nonumber\\
&-&\frac{\theta}{2} \bar{n}(n_1+m_1+2)\rho_{23}(t),\nonumber\\
\frac{\rho_{24}}{dt}&=&-\frac{\theta}{2}(\bar{n}+1) [(2n_1+2m_1+3)\rho_{24}(t)]\nonumber\\
&-&\frac{\theta}{2}\bar{n}[(n_1+1)\rho_{24}(t)-2(m_1+1)\rho_{13}(t)],\nonumber\\
\frac{\rho_{31}}{dt}&=&-\frac{\theta}{2} (\bar{n}+1) [(2n_1+2m_1+1)\rho_{31}(t)-2(n_1+1)\rho_{42}(t)]\nonumber\\
&-&\frac{\theta}{2}(2n_1+m_1+3)\rho_{31}(t),\nonumber\\
\frac{\rho_{32}}{dt}&=&-\theta (\bar{n}+1)(n_1+m_1+1)\rho_{32}(t)\nonumber\\
&-&\frac{\theta}{2} \bar{n}(n_1+m_1+2)\rho_{32}(t),\nonumber\\
\frac{\rho_{33}}{dt}&=&-\theta(\bar{n}+1)[(n_1+m_1+1)\rho_{33}(t)-(m_1+1)\rho_{44}(t)]\nonumber\\
&-&\theta\bar{n}[(m_1+1)\rho_{33}(t)-(n_1+1)\rho_{11}(t)], \nonumber\\
\frac{\rho_{34}}{dt}&=&-\frac{\theta}{2}(\bar{n}+1)[(2n_1+2m_1+3)\rho_{34}(t)]\nonumber\\
&-&\frac{\theta}{2}\bar{n}[(m_1+1)\rho_{34}(t)-2(n_1+1)\rho_{12}(t)],\nonumber\\
\frac{\rho_{41}}{dt}&=&-\theta(n_1+1)(n_1+m_1+1)\rho_{41}(t)\nonumber\\
&-&\frac{\theta}{2}\bar{n}(n_1+m_1+2)\rho_{41}(t),\nonumber\\
\frac{\rho_{42}}{dt}&=&-\frac{\theta}{2}(\bar{n}+1) [(2n_1+2m_1+3)\rho_{42}(t)]\nonumber\\
&-&\frac{\theta}{2}\bar{n}[(n_1+1)\rho_{42}(t)-2(m_1+1)\rho_{31}(t)],\nonumber\\
\end{eqnarray}
\begin{eqnarray}\label{yy}
\frac{\rho_{43}}{dt}&=&-\frac{\theta}{2}(\bar{n}+1)[(2n_1+2m_1+3)\rho_{43}(t)]\nonumber\\
&-&\frac{\theta}{2}\bar{n}[(m_1+1)\rho_{43}(t)-2(n_1+1)\rho_{21}(t)],\nonumber\\
\frac{\rho_{44}}{dt}&=& -\theta(\bar{n}+1)(n_1+m_1+2)\rho_{44}(t)\nonumber\\
&+&\theta\bar{n}[(n_1+1)\rho_{22}(t)+(m_1+1)\rho_{33}(t)].
\end{eqnarray}

 Here $\theta=\gamma_M$ in Markovian case and $\theta=\Gamma(t)$ in non-Markovian case. As we have already pointed out, if we put $n_1=m_1=0$ then one can obtain exactly the same results in \cite{29}. But in our case if  $n \neq  0$ and $m\neq 0$, then the solutions of  motion's equations in Eq. (\ref{yy}) becomes too complicated. Consequently we suppose that $n_1=m_1$ and $\bar{n}=0$. In this case one  obtain the solutions of motion's equations of the density matrix elements for vacuum reservoir in Appendix $B$.

\section*{Appendix B.  Solutions of equations of motion of the density matrix elements for vacuum reservoir in Markovian and non-Markovian cases}
\begin{eqnarray}
\rho_{11}(t)&=& [\rho_{11}(0)+(1+m_1)(\rho_{22}(0)+\rho_{33}(0))+(1+m_1)^2 \rho_{44}(0) ] e^{-2\theta m_1 t}
\nonumber\\
&+&[(1+m_1)^2 \rho_{44}(0)-(1+m_1)(\rho_{22}(0)+\rho_{33}(0)+2(1+m_1)\rho_{44})]e^{-\theta (1+2m_1)t},\nonumber\\
\rho_{12}(t)&=&-(1+m_1)\rho_{34}(0) e^{-\theta/2(3+4m_1)t}+(\rho_{12}(0)+(1+m_1)\rho_{34}(0))e^{-\theta/2(1+4m_1)t},\nonumber\\
\rho_{13}(t)&=&-(1+m_1) \rho_{24}(0) e^{-\theta/2(3+4m_1)t}+(\rho_{13}(0)+(1+m_1)\rho_{24}(0))e^{-\theta/2(1+4m_1)t},\nonumber\\
\rho_{14}(t)&=& \rho_{14}(0)e^{-2 \theta (1+m_1) t},\nonumber\\
\textbf{}\rho_{22}(t)&=&-(1+m_1) \rho_{44}(0) e^{-\theta (1+2m_1)t}+(\rho_{22}(0)+(1+m_1)\rho_{44}(0)) e^{-\theta (1+2m_1)t},\nonumber\\
\rho_{23}(t)&=& \rho_{23}(0) e^{-\theta (1+2m_1)t},\nonumber\\
\rho_{24}(t)&=& \rho_{24}(0) e^{-\theta/2 (3+4m_1)t},\nonumber\\
\rho_{33}(t)&=&[m_1\rho_{44}(0)+\rho_{33}(0)] e^{-\theta (1+2m_1)t},\nonumber \\
\rho_{34}(t)&=& \rho_{34}(0) e^{-\theta/2 (3+4m_1)t},\nonumber\\
\rho_{44}(t)&=& 1-\rho_{11}-\rho_{22}-\rho_{33},\nonumber\\
\rho_{12}(t)&=&\rho_{21}(t),\quad\rho_{13}(t)=\rho_{31}(t), \quad \rho_{32}(t)=\rho_{23}(t)\nonumber\\
\rho_{14}(t)&=&\rho_{41}(t),\quad\rho_{42}(t)=\rho_{24}(t), \quad \rho_{34}(t)=\rho_{43}(t).
\end{eqnarray}

\section*{Acknowledgment}

K. El Anouz acknowledges financial support for this research from the "Centre National pour la Recherche Scientique et Technique" CNRST, Morocco. A. El Allati acknowledges the hospitality of the Abdus Salam
International Center for Theoretical Physics (Trieste, Italy)

\end{document}